\documentclass[]{aa}  

\usepackage{graphicx}
\usepackage{txfonts}
\usepackage{amsmath,dsfont,mathrsfs,mathtools,amssymb}
\usepackage{xcolor}  
\usepackage{empheq}
\usepackage{multirow}
\usepackage{amsfonts}
\usepackage[bbgreekl]{mathbbol}
\usepackage{cancel}
\DeclareSymbolFontAlphabet{\mathbb}{AMSb}
\newcommand{\nuO}{\nu_{0}}
\newcommand{\dr}{\partial}
\newcommand{\bd}{\boldsymbol}
\newcommand{\vvec}{\bd{\varv}}
\newcommand{\avec}{\bd{a}}
\newcommand{\rvec}{\bd{r}}
\newcommand{\nvecO}{\bd{n}_{0}}
\newcommand{\Em}{\mathscr{E}}
\newcommand{\EO}{E_{0}}
\newcommand{\EOnuO}{E_{\nuO}}
\newcommand{\FO}{F_{0}}
\newcommand{\Fvec}{\bd{F}}
\newcommand{\FvecO}{\bd{F}_{0}}
\newcommand{\FvecOnuO}{\bd{F}_{\nuO}}
\newcommand{\PtensO}{\mathbb{P}_{0}}
\newcommand{\gvec}{\bd{g}}
\newcommand{\GOO}{G_{0}^{0}}
\newcommand{\GvecO}{\bd{G}_{0}}
\newcommand{\LO}{L_{0}}
\newcommand{\IO}{I_{0}}
\newcommand{\IOlong}{\IO\left(\rvec, t; \nvecO, \nuO \right)}
\newcommand{\chiO}{\chi_{0}}
\newcommand{\chiOlong}{\chiO\left(\rvec, t; \nvecO, \nuO \right)}
\newcommand{\kappaOth}{\kappa^{\rm th}_{0}}
\newcommand{\kappaOthlong}{\kappa^{\rm th}_{0}\left(\rvec, t; \nvecO, \nuO \right)}
\newcommand{\kappaOthnuO}{\kappa^{\rm th}_{\nuO}}
\newcommand{\kappaOsca}{\kappa^{\rm sca}_{0}}
\newcommand{\kappaOscalong}{\kappa^{\rm sca}_{0}\left(\rvec, t; \nvecO, \nuO \right)}
\newcommand{\etaO}{\eta_{0}\left(\rvec, t; \nvecO, \nuO \right)}
\newcommand{\etaOth}{\eta^{\rm th}_{0}}
\newcommand{\etaOthlong}{\eta^{\rm th}_{0}\left(\rvec, t; \nvecO, \nuO \right)}
\newcommand{\etaOsca}{\eta^{\rm sca}_{0}}
\newcommand{\etaOscalong}{\eta^{\rm sca}_{0}\left(\rvec, t; \nvecO, \nuO \right)}
\newcommand{\OmegaO}{\Omega_{0}}
\newcommand{\kOE}{\kappa_{0E}}
\newcommand{\kOP}{\kappa_{0P}}
\newcommand{\koP}{k_{0P}}
\newcommand{\chinuO}{\chi_{\nuO}}
\newcommand{\chiOF}{\chi_{0F}}
\newcommand{\chiOFi}{\chi^{(i)}_{0F}}
\newcommand{\chiOFiott}{\chi^{(i),\,i=1,3}_{0F}}
\newcommand{\chiOFone}{\chi^{(1)}_{0F}}
\newcommand{\chiOFtwo}{\chi^{(2)}_{0F}}
\newcommand{\chiOFthree}{\chi^{(3)}_{0F}}
\newcommand{\chiOFtens}{\bbchi_{0F}}
\newcommand{\chiOR}{\chi_{0R}}
\newcommand{\koR}{k_{0R}}
\newcommand{\BnuO}{B_{\nuO}}
\newcommand{\BO}{B_{0}}
\newcommand{\Bvec}{\bd{B}}
\newcommand{\lambdap}{\lambda_{p}}
\newcommand{\tr}{t_{r}}
\newcommand{\tf}{t_{f}}
\newcommand{\vf}{\varv_{f}}
\newcommand{\td}{t_{d}}
\newcommand{\vd}{\varv_{d}}
\newcommand{\visctens}{\bbtau_{\text{visc}}}
\newcommand{\Fcond}{\bd{F}_{c}}
\newcommand{\kB}{k_{\rm B}}
\newcommand{\mH}{m_{\rm H}}
\newcommand{\Jvec}{\bd{J}}
\newcommand{\dtn}{\Delta t^{n}}
\newcommand{\rhon}{\rho^{n}}
\newcommand{\rhonpI}{\rho^{n+1}}
\newcommand{\rhostar}{\rho^{\ast}}
\newcommand{\vn}{\vvec^{n}}
\newcommand{\vnpI}{\vvec^{n+1}}
\newcommand{\vstar}{\vvec^{\ast}}
\newcommand{\pn}{p^{n}}
\newcommand{\pnpI}{p^{n+1}}
\newcommand{\pstar}{p^{\ast}}
\newcommand{\En}{E^{n}}
\newcommand{\EnpI}{E^{n+1}}
\newcommand{\Tn}{T^{n}}
\newcommand{\TnpI}{T^{n+1}}
\newcommand{\Tstar}{T^{\ast}}
\newcommand{\Emn}{\Em^{n}}
\newcommand{\EmnpI}{\Em^{n+1}}
\newcommand{\Fn}{\Fvec^{n}}
\newcommand{\FnpI}{\Fvec^{n+1}}
\newcommand{\cV}{c_{V}}
\newcommand{\LnpI}{L^{n+1}}
\newcommand{\kPstar}{k_P^{\ast}}
\newcommand{\kPstarijk}{k^{\ast}_{P\,ijk}}
\newcommand{\kRstar}{k_R^{\ast}}
\newcommand{\kRnpI}{k_R^{n+1}}
\newcommand{\Rstar}{R^{\ast}}
\newcommand{\RnpI}{R^{n+1}}
\newcommand{\Lstar}{L^{\ast}}
\newcommand{\Kstar}{K^{\ast}}
\newcommand{\hijk}{h_{ijk}}

\begin{document}

   \title{Non-LTE radiation hydrodynamics in PLUTO}
   \author{S. Colombo \inst{1,} \inst{2,} \inst{3}
          \and
          L. Ibgui\inst{2}
          \and
          S. Orlando\inst{3}
          \and
          R. Rodr{\'i}guez\inst{4}
          \and
          G. Espinosa\inst{4}
          \and 
          M. Gonz{\'a}lez\inst{5}
          \and
          C. Stehl{\'e}\inst{2}
          \and 
          G. Peres\inst{1}
          }

   \institute{Dipartimento di Fisica e Chimica,                     Università degli Studi di Palermo,
              via Archirafi 36, Palermo, Italy\\
              \email{salvatore.colombo@inaf.it}
         \and
             LERMA, Observatoire de Paris, Sorbonne Universit{\'e}, Universit{\'e} de
             Cergy-Pontoise, CNRS, Paris,France;
          \and 
             INAF - Osservatorio Astronomico di Palermo, Palermo, Italy;
          \and
             Universidad de Las Palmas de Gran Canaria, Gran Canaria, Spain;
          \and 
             Universit{\'e} Paris Diderot, Sorbonne Paris Cit{\'e}, AIM, UMR 7158, CEA, 91191, Gif-sur-Yvette, France}

   \date{}

 
  \abstract
   {Modeling the dynamics of most astrophysical structures requires an adequate description of the radiation-matter interaction.
   Several numerical (magneto)hydrodynamics codes were upgraded with a radiation module to fulfill this request.
   However, those among them that use either the flux-limited diffusion (FLD) or the M1 radiation moment approaches
   are restricted to the local thermodynamic equilibrium (LTE).
   This assumption may be not valid in some astrophysical cases.}
   {We present an upgraded version of the LTE radiation-hydrodynamics module implemented in the PLUTO code, originally developed by \cite{Kolb_et_al_2013},
   which we have extended to handle non-LTE regimes.}
   {Starting from the general frequency-integrated comoving-frame equations of radiation hydrodynamics (RHD),
   we have justified all the assumptions made to obtain the non-LTE equations actually implemented in the module,
   under the FLD approximation.
   An operator-split method is employed, with two substeps: the hydrodynamic part is solved with an explicit method by the solvers
   already available in PLUTO, the non-LTE radiation diffusion and energy exchange part is solved with an implicit method.
   The module is implemented in the PLUTO environment. It uses databases of radiative quantities that can be provided independently by the user:
   the radiative power loss, the Planck and Rosseland mean opacities. In our case, these quantities were determined from
   a collisional-radiative steady-state model (CRSS), and tabulated as functions of temperature and density.}
   {Our implementation has been validated through different tests, in particular radiative shock tests.
   The agreement with the semi-analytical solutions (when available) is good, with a maximum error of $7\%$.
   Moreover, we have proved that non-LTE approach is of paramount importance to properly model accretion shock structures.}
   {Our radiation FLD module represents a step toward the general non-LTE RHD modeling.
   The module is available, under request, for the community.}
\keywords{Radiation -- Radiation Hydrodynamics -- NLTE -- PLUTO code}
\maketitle

\section{Introduction}  \label{sec:introduction}

The description of fluid motion in many astrophysical systems requires to consider the effects of radiation through
its momentum and energy exchanges with matter, which are described by the radiation hydrodynamics (RHD) equations.
Examples of relevant systems include, to name just a few, star formation structures (e.g., \citealt{Krumholz_et_al_2009,Commercon_et_al_2011b,
Vaytet_et_al_2012,Vaytet_et_al_2013a,Davis_et_al_2014, Skinner_and_Ostriker_2015, Raskutti_et_al_2016}), protoplanetary discs (e.g., \citealt{Flock_et_al_2013,Flock_et_al_2017}),
accretion flows around young stellar objects such as T~Tauri stars (\citealt{Costa_et_al_2017,Colombo_et_al_2019_2,de_Sa_et_al_inprep}),
accretion around black holes (e.g., \citealt{Hirose_et_al_2009,Jiang_et_al_2014}).
In all these cases, radiation is coupled with matter from the dynamical and energetic points of view.

A direct coupling of the radiation effects to the hydrodynamic equations (RHD) or even the magnetohydrodynamic equations (RMHD)
requires to solve the radiative transfer equation (RTE) and, in non-local thermodynamic equilibrium (non-LTE or NLTE) regimes,
the kinetic equilibrium equations, at each time step, in order to infer the radiation four-force density vector
that describes the momentum and energy coupling between radiation and matter. However, this is a challenge that is, still to date,
far beyond the current capabilities of computers. The main reason is that the RTE is an integro-differential equation, whose unknown,
the specific intensity, depends on seven variables in a three-dimensional description
(e.g., \citealt{pomraning_book_1973, Mihalas_and_Mihalas_book_1984, Castor_book_2004, Hubeny_and_Mihalas_book_2014}).
Fully solving the 3D radiative transfer equation itself requires a dedicated approach (e.g., \citealt{Ibgui_et_al_2013}) that can be
used for the generation of synthetic spectra, without coupling to hydrodynamic evolution.

A work around is to solve directly the equations that involve the radiation moments. These are obtained from angular moments of the RTE.
Such an approach entails the creation of a hierarchy of moments, and necessitates the use of closure relations.
In fact, the equations describing the conservation of mass, matter momentum, and matter energy, are solved together
with the equations describing the conservation of radiation momentum and radiation energy.

The most accurate of the radiation moment techniques is the variable Eddington tensor (VET) method
\citep{Stone_et_al_1992,Gehmeyr_and_Mihalas_1994,Gnedin_and_Abel_2001,Hayes_and_Norman_2003,Hubeny_and_Burrows_2007,Jiang_et_al_2012}.
The method has been applied to the total, or frequency-integrated, radiation moment equations.
The Eddington tensor is defined as the ratio of the radiation pressure to the radiation energy.
The method consists in solving the moment equations, while assuming that the Eddington tensor is known.
Then, once the structure of the medium is determined, the Eddington tensor is updated
by solving the RTE with the short-characteristics method \citep{Davis_et_al_2012}.
Despite its good precision, this method is very costly from the computational point of view.

Another technique, less accurate than the VET method, but much cheaper from the computational point of view,
is the so-called M1 approximation \citep{Levermore_1984,Dubroca_and_Feugeas_1999}.
The difference with the VET method, for the total radiation moment equations, is that the Eddington tensor is
provided by an analytical closure relation; we do not solve the RTE.
The M1 method has been implemented in several multidimensional RHD or RMHD codes, not only in the frequency-integrated approach
(HERACLES: \citealt{Gonzalez_et_al_2007}, ATON: \citealt{Aubert_and_Teyssier_2008},
RAMSES: \citealt{Rosdahl_et_al_2013}, ATHENA: \citealt{Skinner_and_Ostriker_2013},  PLUTO: \citealt{Melon_Fuksman_and_Mignone_2019}),
but also in the multigroup approach (HERACLES: \citealt{Vaytet_et_al_2013b}, FORNAX: \citealt{Skinner_et_al_2019}).

Finally, we mention the third, radiation moment based, technique: the Flux-limited Diffusion approximation (FLD)
\citep{Alme_and_Wilson_1973,Levermore_and_Pomraning_1981}. Though less accurate than M1, especially in the optically thin regions,
it is the most widely used method in the R(M)HD codes, because it is the simplest one, the most robust, the most efficient in terms
of computational cost, and provides very good results in flows where optically thick regions are of paramount importance.
The difference with M1 is that the radiation flux equation is not solved; instead, the comoving-frame (also known as the fluid frame;
i.e., the frame moving with the macroscopic velocity of the fluid) radiation flux is determined,
through a Fick's law, as a quantity that is antiparallel to the gradient of the radiation energy density, and that recovers,
thanks to an ad-hoc function (the flux-limiter), the correct relations between flux and energy in the two asymptotic regimes:
free-streaming and (static or dynamic) diffusion.
The FLD method has been implemented in many multidimensional RHD or RMHD codes in astrophysical context, in the
frequency-integrated approach (ZEUS: \citealt{Turner_and_Stone_2001},
ORION: \citealt{Krumholz_et_al_2007}, V2D: \citealt{Swesty_and_Myra_2009}, NIRVANA: \citealt{Kley_et_al_2009},
RAMSES: \citealt{Commercon_et_al_2011a}, CASTRO: \citealt{Zhang_et_al_2011}, PLUTO: \citealt{Kolb_et_al_2013,Flock_et_al_2013}),
and in the multigroup approach (CRASH: \citealt{van_der_Holst_et_al_2011}, CASTRO: \citealt{Zhang_et_al_2013},
FLASH: \citealt{Klassen_et_al_2014}, RAMSES: \citealt{Gonzalez_et_al_2015}).

All the above R(M)HD codes that use the M1 or the FLD approximations are restricted to the Local Thermodynamic Equilibrium (LTE) approximation.
While the LTE regime is a good approximation in the optically thick parts of a flow where densities are high enough,
it can no longer be advocated in optically thin parts, such as in the post-shock region of an accretion column (e.g., \citealt{Ardila_2007}).
It is therefore important to get free from the restrained LTE assumption, and to develop more general RHD algorithms,
that would be capable of handling a non-LTE regime, while the LTE regime could just be a particular case to which the non-LTE regime
would tend in parts of a flow that would have appropriate density and temperature conditions.
In this work, we present an extended version of the radiation module for PLUTO, originally developed by \cite{Kolb_et_al_2013}
in the LTE approximation, using the FLD method with the frequency-integrated comoving-frame radiation quantities.
We have expanded the capabilities of this module, so that it can now allow for non-LTE regimes.
We have adopted an implicit scheme to couple the gas and radiation energy exchange, after linearization of a non-LTE radiative emission function,
as implemented for a line emission function in ORION \citep{Cunningham_et_al_2011,Krumholz_et_al_2011,Krumholz_et_al_2012}.

The paper is structured as follows: In Sect.~\ref{sec:the_equations}, we introduce the full system of equations to
describe radiation hydrodynamic. We justify all the assumptions used in this work, and we present the approximated equations solved by the code.
In Sect.~\ref{sec:opacity}, we explain the theoretical model used to generate the opacity and the radiative power loss databases used by our module.
In Sect.~\ref{sec:implementation}, we explain the numerical implementation used in the code.
In Sect.~\ref{sec:test}, we validate the modifications made in the code through some test cases.
In Sect.~\ref{sec:LTE_vs_NLTE_shocks}, we compare the two radiative shock structures obtained by letting evolve
a given flow under, respectively, forced LTE regime, and non-LTE regime. Finally, in Sect.~\ref{sec:conclusion}, we draw our conclusions.

\section{The equations} \label{sec:the_equations}

The originality of our radiation module coupled to the three-dimensional (3D) MHD code PLUTO lies in the fact that it considers the non-LTE radiation regime:
we have removed the LTE assumption that is commonly adopted in radiation hydrodynamics implementations. A bunch of RHD modules and codes exist and are explained
in the literature, with a wide variety of approximations and numerical models. It has, therefore, appeared useful to us to start from the basic general physical
RHD equations, formulated in the comoving frame,
then to clarify and list the approximations that can be done in order to end up with the simplified equations that are actually solved by radiation modules. 
Our purpose is to start from the most general RHD equations and to introduce, successively, and step by step, all the simplifying assumptions that we have considered,
until we obtain the RHD equations that are actually solved by our module and coupled to the MHD equations solved by PLUTO: these are synthesized in the next section
(\S~\ref{subsec:PLUTO_RMHD_equations}). We also discuss each of these assumptions and their consequence on the RHD modeling of a flow.
\citet{Kolb_et_al_2013} have chosen to express the radiation quantities in the comoving frame.
We have followed their choice of reference frame. 

\subsection{The general comoving-frame RHD equations} \label{subsec:general_RHD_equations}
We start with the general comoving-frame frequency-integrated RHD equations, for a nonrelativistic flow,
a nonviscous and nonconducting fluid, but subject to external gravity, and without nuclear reactions.
In the whole paper, and for all applications of our module, we will always restrain to such flows.
The equations are, to $O\left(\varv/c\right)$ (e.g., \citealt{Mihalas_and_Mihalas_book_1984}):
\begin{subequations}
   \begin{alignat}{2}
      &\frac{\dr \rho}{\dr t} + \nabla \cdot \left( \rho \, \vvec\right)        = 0 \:,    \label{eq:continuity}     \\
      &\frac{\dr}{\dr t}\left( \rho \, \vvec \right) +  \nabla \cdot \left( \rho \, \vvec \otimes \vvec + p \, \mathbb{I} \right) = \rho \, \gvec + \GvecO    \label{eq:matter_momentum} \:, \\
      &\frac{\dr}{\dr t} \left( \Em + \tfrac{1}{2} \, \rho \, \varv^2 \right) +  \nabla \cdot \left[ \left( \Em + \tfrac{1}{2} \, \rho \, \varv^2 + p \right) \, \vvec  \right] = \rho \, \gvec \cdot \vvec \label{eq:matter_energy} \\ 
      & \hspace{6cm} 
      + \GvecO \cdot \vvec + \, c \, \GOO. \nonumber
   \end{alignat}
   \begin{equation} \label{eq:Erad_integrated}
      \left\{
      \begin{alignedat}{2}
      &\frac{\dr \EO}{\dr t} + \nabla \cdot \FvecO  + \vvec \cdot \frac{\dr}{\dr t} \left(\frac{\FvecO}{c^{2}}\right) + \nabla \cdot \left(\EO \, \vvec \right) + \PtensO : \nabla \, \vvec \\
      &+ 2 \, \left(\frac{\FvecO}{c^{2}}\right)  \cdot \avec 
      = \, - c \, \GOO \:,
      \end{alignedat}
      \right.
   \end{equation}
   \begin{equation} \label{eq:Frad_integrated}
      \left\{
      \begin{alignedat}{2}
      &\frac{\dr}{\dr t} \left(\frac{\FvecO}{c^{2}} \right) + \nabla \cdot \, \PtensO + \left( \vvec \cdot \nabla \right) \left(\frac{\FvecO}{c^{2}} \right) + \left(\frac{\FvecO}{c^{2}} \cdot \nabla \right) \vvec \\
      &+ \left( \nabla \cdot \, \vvec \right) \left(\frac{\FvecO}{c^{2}}\right)
       + \frac{\dr \PtensO}{\dr t} \, \frac{\vvec}{c^{2}}
      + \frac{1}{c^{2}} \left( \PtensO \, \avec + \EO \, \avec \right) \, = \, - \, \GvecO  \:,
      \end{alignedat}
      \right.
   \end{equation}
\end{subequations}
Equations (\ref{eq:continuity}), (\ref{eq:matter_momentum}), and (\ref{eq:matter_energy})
respectively describe the conservation of mass (continuity equation), matter momentum, and matter total energy.
Equations (\ref{eq:Erad_integrated}) and (\ref{eq:Frad_integrated}) respectively describe the total, or frequency integrated,
radiation energy conservation and radiation momentum conservation.

We use the following notations: $\rho$ is the mass density, $\vvec$ is the fluid (macroscopic) velocity,
$a \equiv \dr \vvec / \dr t$ is the local fluid acceleration, $p$ is the matter pressure, $\gvec$ is the external gravity, $\Em$ is the matter internal energy density.
In our approach, all radiation quantities are considered in the comoving frame. Following \citet{Mihalas_and_Mihalas_book_1984},
we denote them with a subscript ``0''. $\GOO$ and $\GvecO$ are, respectively, the time component and the space components of the radiation four-force density vector 
$G^{\alpha}_0 \equiv \left(\GOO,\GvecO\right)$,
and are defined by \citep{Mihalas_and_Mihalas_book_1984}:
\begin{subequations} \label{eq:rad_4_force} 
\begin{alignat}{1}
   &\GOO\left(\rvec, t \right) \equiv 
   \frac{1}{c} \int_{0}^{\infty} d\nuO \oint_{4\pi} \, \left( \chi_{0} \, I_{0} - \eta_{0} \right) \, d\OmegaO \:, \label{eq:G00}\\ 
   &\GvecO \left(\rvec, t \right) \equiv \frac{1}{c} \int_{0}^{\infty} d\nuO \oint_{4\pi} \, \left( \chi_{0} \, I_{0} - \eta_{0} \right) \, \nvecO \, d\OmegaO \:, \label{eq:Gvec0}
\end{alignat}
\end{subequations}
where $\IOlong$ is the comoving-frame specific intensity defined at position $\rvec$, time $t$, for the comoving-frame direction unit vector $\nvecO$
and frequency $\nuO$, and where $\chiOlong$ and $\etaO$ are, respectively,
the extinction coefficient (${\rm cm^{-1}}$) and emission coefficient or emissivity ($ {\rm erg}\;{\rm cm^{-3}}\;{\rm s^{-1}}\;{\rm sr^{-1}}\;{\rm Hz^{-1}})$:
\begin{subequations} \label{eq:chi0_eta0} 
\begin{alignat}{2}
   &\chiOlong = \kappaOthlong + \kappaOscalong \:, \label{eq:chi0} \\
   &\etaO = \etaOthlong + \etaOscalong \:, \label{eq:eta0} 
\end{alignat}
\end{subequations}
where $\kappaOthlong$, $\kappaOscalong$, $\etaOthlong$, $\etaOscalong$ are, respectively,  the comoving-frame thermal absorption coefficient,
scattering coefficient, thermal emission coefficient, and scattering emission coefficient.

Also, $\EO$, $\FvecO$, and $\PtensO$ are the total, comoving-frame, energy density, radiation flux, and radiation pressure:
\begin{subequations} \label{eq:total_radiation_moments} 
\begin{alignat}{2}
   &\EO\left(\rvec, t \right) \, \equiv \, \frac{1}{c} \, \int_{0}^{\infty} \, \oint_{4\pi} \IOlong \, d\OmegaO \, d\nuO \:,                                \label{eq:EO} \\
   &\FvecO\left(\rvec, t \right) \, \equiv \, \int_{0}^{\infty} \, \oint_{4\pi} \IOlong \, \nvecO \, d\OmegaO \, d\nuO \:,                                  \label{eq:FO} \\
   &\PtensO\left(\rvec, t \right) \, \equiv \, \frac{1}{c} \, \int_{0}^{\infty} \,  \oint_{4\pi} \IOlong \, \nvecO \otimes \nvecO \, d\OmegaO \, d\nuO  \:. \label{eq:PO}
\end{alignat}
\end{subequations}

The radiation moment approach couples the three matter-related conservation equations (\ref{eq:continuity}), (\ref{eq:matter_momentum}), (\ref{eq:matter_energy}),
an equation of state (EOS) for matter, $p=p(\rho,T)$, an equation for the matter internal energy,
or caloric equation of state, $\Em=\Em(\rho,T)$, and the two radiation moment equations (\ref{eq:Erad_integrated}) and (\ref{eq:Frad_integrated}).
This system needs to be closed by relations that link the three radiation moments (cf. \S~\ref{subsubsec:approx_4_FLD}).

\subsection{The non-LTE approximate RHD equations} \label{subsec:NLTE_approx_RHD}

\subsubsection{The approximate radiation four-force density} \label{subsubsec:radiation_four_force}

Our objective is to determine the comoving-frame radiation four-force density vector $\left(\GOO,\GvecO\right)$ (cf. Eq.~(\ref{eq:G00}) and (\ref{eq:Gvec0})),
without having to calculate the specific intensity.

We focus on $\GOO$. We assume coherent and isotropic scattering in the comoving frame.
Then, one can show \citep{Mihalas_and_Auer_2001} that  
\begin{equation}  \label{eq:coherent_isotropic_scattering_eta_sca}
  \etaOsca(\rvec, t;\nuO) \, = \, \kappaOsca(\rvec, t;\nuO) \, \oint_{4\pi}  \, \IO(\rvec, t; \nvecO',\nuO) \, \frac{d\OmegaO'}{4 \pi}
\end{equation}
Then, the scattering terms in Eq.~(\ref{eq:G00}) cancel out.
We also assume isotropic thermal absorption $\kappaOth(\rvec, t;\nuO)$.
Finally, we obtain:
\begin{equation} \label{eq:G00_simplified_integrated}
   c \, \GOO \, = \, c \, \kOE \, \EO - \LO
\end{equation}
where $\LO$ is the total radiation emission per unit volume and time (or radiative power loss):
\begin{equation} \label{eq:L0}
    \LO\left(\rvec, t \right) \, \equiv \, \int_{0}^{\infty} d\nuO \oint_{4\pi} \etaOth(\rvec, t; \nvecO,\nuO) \, d\OmegaO
\end{equation}
and where $\kOE$ is the energy-weighted (or absorption) mean opacity, which is defined in the comoving frame as follows:
\begin{equation} \label{eq:k0E}
    \kOE  \, \equiv \, \frac{\int_{0}^{\infty} \kappaOthnuO \, \EOnuO \, d\nuO}{\int_{0}^{\infty} \, \EOnuO \, d\nuO}
\end{equation}
where we use the simplified notation $\kappaOthnuO = \kappaOth(\rvec, t;\nuO)$.

We focus on $\GvecO$. We assume the additional assumption of isotropic thermal emission $\etaOth(\rvec, t; \nuO)$.
Then, the emission terms vanish in Eq.~(\ref{eq:Gvec0}). We end up with:
\begin{equation}  \label{eq:Gvec0_simplified_integrated}
      \GvecO    \, = \, \frac{1}{c} \, \chiOFtens \, \FvecO
\end{equation}
where $\chiOFtens$ is the flux-weighted (or radiation momentum) mean opacity, defined in the comoving frame, as follows \citep{Mihalas_and_Auer_2001}:
\begin{subequations} 
  \begin{empheq}[left={\empheqlbrace}]{alignat=4}
      &\chiOFtens \, \FvecO \, = \int_{0}^{\infty} \chinuO \, \FvecO(\nuO) \, d\nuO \label{eq:chiFO_notation}\\
      &\text{with} \nonumber \\
      &\chiOFtens \, = \,
   \left(
   \begin{matrix}
       \chiOFone  & 0         & 0 \\
           0      & \chiOFtwo & 0 \\ 
           0      &      0    & \chiOFthree 
   \end{matrix}
   \right) \label{eq:chiFO_matrix}
  \end{empheq}
\end{subequations}
where we use the simplified notation $\chinuO = \chi_{0}\left(\rvec, t;\nuO \right)$.

\subsubsection{The approximate mean opacities} \label{subsubsec:approx_mean_opacities}

The mean opacities $\kOE$ (\ref{eq:k0E}) and $\chiOFtens$ (\ref{eq:chiFO_notation}) are not known in advance,
since they depend on the unknown radiation energy $\EOnuO$ and radiation flux $\FvecOnuO$.
Our objective is to provide approximate expressions for these quantities.\\

If we consider an optically thick medium in the equilibrium diffusion regime, we know that \citep{Mihalas_and_Mihalas_book_1984}:
\begin{subequations}
  \begin{empheq}[left={\empheqlbrace}]{alignat=4}
      &\EOnuO    \, & &= \, & & \frac{4\,\pi}{c} \, \BnuO                     \label{eq:E0nu0_equilibrium_diffusion}\\
      &\FvecOnuO \, & &= \, & &-\frac{4\,\pi}{3\, \chinuO} \, \nabla \, \BnuO \label{eq:F0nu0_equilibrium_diffusion}
  \end{empheq}
\end{subequations}
where $\BnuO = B\left(\nuO,T_{0}\right)$ is the Planck function at material temperature $T_{0}$. As a consequence, using Eq. (\ref{eq:k0E}):
\begin{equation} \label{eq:k0E_eq_k0P}
      \kOE  \, = \, \kOP
\end{equation}
where $\kOP$ is the Planck mean opacity, defined as:
\begin{equation} \label{eq:k0P}
   \kOP  \, \equiv \, \frac{\int_{0}^{\infty}    \kappaOthnuO \, \BnuO \, d\nuO}{\int_{0}^{\infty}    \, \BnuO \, d\nuO}
\end{equation}

\noindent From Eq.~(\ref{eq:F0nu0_equilibrium_diffusion}), we find, after frequency integration, that:
\begin{equation}  \label{eq:F0_equilibrium_diffusion}
    \FvecO  \, = \, - \frac{4\pi}{3\,\chiOR}  \nabla \BO
\end{equation}
where $\BO = c a_{R} T_{0}^{4} / 4 \pi$ is the frequency-integrated Planck function ($a_{R}$: radiation density constant),
and where $\chiOR$ is the Rosseland mean opacity, defined as:
\begin{equation}  \label{eq:chiOR}
      \frac{1}{\chiOR}  \, \equiv \, \frac{\int_{0}^{\infty}    \frac{1}{\chinuO} \frac{\dr \BnuO}{\dr T} \, d\nuO}{\int_{0}^{\infty}    \, \frac{\dr \BnuO}{\dr T} \, d\nuO}
\end{equation}
In the equilibrium diffusion regime, relations (\ref{eq:G00_simplified_integrated}) and (\ref{eq:Gvec0_simplified_integrated})
can be replaced with:
\begin{subequations}
  \begin{empheq}[left={\empheqlbrace}]{alignat=4}
      &c \, \GOO  \, & &= \, & &c \, \kOP   \, \EO - \LO          \label{eq:G00_simplified_integrated_equilibrium_diffusion} \\
      &\GvecO     \, & &= \, & &\frac{1}{c} \, \chiOR  \, \FvecO  \label{eq:Gvec0_simplified_integrated_equilibrium_diffusion}
  \end{empheq}
\end{subequations}

Let us consider now the optically thin regime. In that case, $F_{\nuO} \to c E_{\nuO}$
\citep{Mihalas_and_Mihalas_book_1984,Krumholz_et_al_2007}. Then, we have $\chiOFiott  \to  (\kOE \text{ + a scattering term})$.
If, for example, we consider Thomson scattering by free electrons of density $n_{e}$, with (frequency independent) cross section $\sigma_{e}$, we have
 $\chiOFiott \to  \kOE  + \sigma_{e}\, n_{e}$.
Now, one can show \citep{Mihalas_and_Mihalas_book_1984} that, in this optically thin regime, and assuming LTE thermal emission (Kirchhoff's law),
the Planck mean $\kOP$ is the appropriate mean to use in order to best approximate $\GOO$. Then, if we neglect scattering,
we could, under the LTE thermal emission assumption, adopt, in optically thin regime, $\chiOFiott=\kOE=\kOP$.
However, this assumption is not physically consistent, because, in an optically thin medium, we are in a non-LTE regime.
And, in our approach, we do not assume LTE \textit{a priori}.
As a result, we choose \textit{not} to set $\chiOFi=\kOP$ in the optically thin regime,
because there is no reason for such a specification to improve the accuracy.\\

In conclusion, first, we approximate $\kOE$ with the Planck mean $\kOP$; second, we consider the flux spectrum
to be the same along each direction, and then replace the tensor $\chiOFtens$ with a scalar $\chiOF$
that we approximate with the Rosseland mean $\chiOR$. We use the approximate expressions
(\ref{eq:G00_simplified_integrated_equilibrium_diffusion}) and (\ref{eq:Gvec0_simplified_integrated_equilibrium_diffusion})
to calculate the components $\GOO$ and $\GvecO$ of the radiation four-force density vector.
This way, we follow the idea of \citet{Krumholz_et_al_2007} of giving accuracy priority to optically thick parts of the flow.
The Planck mean opacity $\kOP$, the Rosseland mean opacity $\chiOR$, and the total radiation emission, or radiative losses, $\LO$,
have been tabulated as functions of density and temperature, for a solar composition of the plasma, assuming a non-LTE regime \citep{Rodriguez_et_al_2018}.
A summary of the adopted physical assumptions is provided in Sect. \ref{sec:opacity} \\

Note that our radiation module \textit{never} assumes \textit{a priori} LTE, but always considers, from the outset,
the non-LTE equations. Now, if, at a given time and position, the properties of the simulated medium are such as the LTE
conditions prevail, then the calculations will provide the same results as one would obtain from LTE equations:
$c \, \GOO \xrightarrow[\text{LTE}]{} \, c \, \kOP  \, \EO  - \kOP \,  4\,\pi \,\BO$.

Within the above approximations, we can rewrite the system (\ref{eq:continuity}) - (\ref{eq:Frad_integrated})
as follows:
\begin{subequations}
   \begin{equation} \label{eq:continuity_integrated_approx}
      \left\{
      \begin{alignedat}{2}
   &\frac{\dr \rho}{\dr t} + \nabla \cdot \left( \rho \, \vvec\right)        = 0 \:,\\
    \end{alignedat}
      \right.
   \end{equation}
   \begin{equation} \label{eq:matter_momentum_integrated_approx}
      \left\{
      \begin{alignedat}{2}
   &\frac{\dr}{\dr t}\left( \rho \, \vvec \right) +  \nabla \cdot \left( \rho \, \vvec \otimes \vvec + p \, \mathbb{I} \right) = \rho \, \gvec + \frac{1}{c} \, \chiOR \, \FvecO \:, \\
    \end{alignedat}
      \right.
   \end{equation}
   \begin{equation} \label{eq:matter_energy_integrated_approx}
      \left\{
      \begin{alignedat}{2}
   &\frac{\dr}{\dr t} \left( \Em + \tfrac{1}{2} \, \rho \, \varv^2 \right) +  \nabla \cdot \left[ \left( \Em + \tfrac{1}{2} \, \rho \, \varv^2 + p \right) \, \vvec  \right] = \rho \, \gvec \cdot \vvec \\
   & \hspace{3cm}
   + \frac{1}{c} \, \chiOR  \, \FvecO \cdot \vvec + \, c \, \kOP \, \EO - \LO\:, \\
    \end{alignedat}
      \right.
   \end{equation}
   \begin{equation} \label{eq:Erad_integrated_approx}
      \left\{
      \begin{alignedat}{2}
   &\frac{\dr \EO}{\dr t} + \nabla \cdot \FvecO  + \vvec \cdot \frac{\dr}{\dr t} \left(\frac{\FvecO}{c^{2}}\right) + \nabla \cdot \left(\EO \, \vvec \right) + \PtensO : \nabla \, \vvec \\
   &+ 2 \, \left(\frac{\FvecO}{c^{2}}\right)  \cdot \avec
      = \, \LO - c \, \kOP   \, \EO \:, \\
    \end{alignedat}
      \right.
   \end{equation}
   \begin{equation} \label{eq:Frad_integrated_approx}
      \left\{
      \begin{alignedat}{2}
   &\frac{\dr}{\dr t} \left(\frac{\FvecO}{c^{2}} \right) + \nabla \cdot \PtensO + \left( \vvec \cdot \nabla \right) \left(\frac{\FvecO}{c^{2}} \right) + \left(\frac{\FvecO}{c^{2}} \cdot \nabla \right) \vvec \\
   &+ \left( \nabla \cdot \, \vvec \right) \left(\frac{\FvecO}{c^{2}}\right)
    + \frac{\dr \PtensO}{\dr t} \, \frac{\vvec}{c^{2}}
    + \frac{1}{c^{2}} \left( \PtensO \, \avec + \EO \, \avec \right)  \\
   &\, = \, - \, \frac{1}{c} \, \chiOR  \, \FvecO.
    \end{alignedat}
      \right.
   \end{equation}
\end{subequations}

\subsubsection{Scaling of terms in the radiation moment equations} \label{subsec:terms_scaling}

The orders of magnitude of the terms involved in the LTE total radiation moment equations have been evaluated
in the comoving-frame equations \citep{Mihalas_and_Mihalas_book_1984,Stone_et_al_1992}, and in the
mixed-frame equations \citep{Mihalas_and_Mihalas_book_1984,Krumholz_et_al_2007,Skinner_and_Ostriker_2013}.
We apply here the same approach to the non-LTE equations.

\paragraph{The three asymptotic physical regimes}\mbox{}\\ \label{paragraph:the_3_asymptotic_regimes}

We summarize the characteristics of the three asymptotic physical regimes, as classified by \citet{Mihalas_and_Mihalas_book_1984}
and used by \citet{Krumholz_et_al_2007} and \citet{Skinner_and_Ostriker_2013}.
We denote by $\ell$ the characteristic structural length at a given position and along a given direction in the flow,
and $\lambdap(\nuO) \equiv 1/\chiO(\nuO)$ the photon mean free path
(distance traveled by a photon before being thermally absorbed or scattered).
The optical depth along the distance $\ell$ in the flow, at frequency $\nuO$, is
$\tau(\nuO) \equiv \chiO(\nuO) \, \ell = \ell/\lambdap(\nuO)$.
We distinguish the free-streaming regime in an optically thin medium ($\tau(\nuO) \ll 1$), where a photon can
move freely in the medium at the speed of light, and the diffusion regime in an optically thick
medium ($\tau(\nuO) \gg 1$), where a photon travels one free path between interactions (thermal absorption or
scattering) with matter.
In the diffusion case, we differentiate two subcases. The first one is the static diffusion regime for media with no
or small enough velocity. Photons, which are trapped in the material, diffuse through a random walk process.
The second one is the dynamic diffusion regime for media with large enough velocity. Photons, which are also
trapped in the material, can be advected by the material motion faster than they can diffuse.
The distinction between these two subcases in an optically thick medium can be quantified by appropriate parameters
\citep{Mihalas_and_Klein_1982,Mihalas_and_Mihalas_book_1984,Krumholz_et_al_2007,Skinner_and_Ostriker_2013}.
We introduce the fluid-flow timescale $\tf \equiv \ell/\vf$, the typical time for a fluid particle to cross distance $\ell$ at characteristic velocity $\vf$,
the diffusion timescale $\td \equiv \ell^{2}/ c \lambdap$, the typical time for a photon to cross distance $\ell$ through a random walk composed of steps of
length $\lambdap$ (mean free path) covered at the speed of light \citep{Mihalas_and_Mihalas_book_1984}. The corresponding characteristic photon diffusion
velocity is $\vd \equiv \ell/\td$. We can then express quantitatively the physical distinction, discussed above, between the static diffusion regime,
$\vd \gg \vf$ (equivalent to $\td \ll \tf$), and the dynamic diffusion regime, $\vf \gg \vd$ (equivalent to $\tf \ll \td$).
We introduce a characteristic $\beta \equiv \vf/c \ll 1$ (nonrelativistic flow).
This way, we can synthesize the quantitative classification of the three asymptotic regimes as follows \citep{Krumholz_et_al_2007}: 
\begin{subequations} \label{eq:the_3_asymptotic_regimes}
  \begin{empheq}[left={\empheqlbrace}]{alignat=10}
      &\text{free streaming:} \quad & &\tau \ll 1 \label{eq:free_streaming_regime}  \\
      &\text{diffusion:}      \quad & &\tau \gg 1  \quad &\text{static:}  \quad & & \beta \, & &\ll \,&\frac{1}{\tau} \label{eq:diffusion_static} \\
      &                             & &                  &\text{dynamic:} \quad & & \beta \, & &\gg \,&\frac{1}{\tau} \label{eq:diffusion_dynamic}
  \end{empheq}
\end{subequations}
where $\tau(\nuO)$ is the monochromatic optical depth as defined above. We extend this classification
to the frequency-integrated optical depth.

\paragraph{Relative sizes of the terms}\mbox{}\\ \label{paragraph:relative_sizes}

We analyse the relative sizes of the terms of the comoving-frame radiation moment equations
(\ref{eq:Erad_integrated_approx}) and (\ref{eq:Frad_integrated_approx}),
by evaluating the order of magnitude of the evolution of each term on the characteristic structural length $\ell$.
We have already introduced, above, the characteristic times for a photon to cross $\ell$: the diffusion timescale $\td$ in a static diffusion regime,
and the fluid-flow timescale $\tf$ in a dynamic diffusion regime. In a free-streaming regime, the characteristic time for a photon to cross $\ell$
is the radiation-flow timescale $\tr \equiv \ell/c$.\\
One can link the three total radiation moments, $\EO$, $\FvecO$, $\PtensO$ in the three asymptotic regimes described above.
One can show that \citep{Mihalas_and_Mihalas_book_1984} in the free streaming regime:
\begin{subequations} \label{eq:E0_F0_P0_streaming_regime}
  \begin{empheq}[left=\text{free streaming\quad}{\empheqlbrace}]{alignat=10}
      &\; \FvecO  & &\xrightarrow[\tau \ll 1]{\beta \ll 1} &\, &c\, \EO \, \nvecO + \bd{O}(\beta) \label{eq:Fvec0_streaming}  \\
      &\; \PtensO & &\xrightarrow[\tau \ll 1]{\beta \ll 1} &\, & \, \EO \, \nvecO \otimes \nvecO + \mathbb{O}(\beta) \label{eq:Ptens0_streaming}
  \end{empheq}
\end{subequations}
One can also show that \citep{Mihalas_and_Mihalas_book_1984} in the, either static of dynamic, first order equilibrium diffusion regime:
\begin{subequations} \label{eq:E0_F0_P0_diffusion_regime}
  \begin{empheq}[left=\text{diffusion (static or dynamic)\quad}{\empheqlbrace}]{alignat=10}
      &\; \FvecO  & &= &\, &- \frac{c}{\chiOR} \nabla \cdot \PtensO \label{eq:Fvec0_equilibrium_diffusion} \\
      &\; \PtensO & &= &\, &  \frac{1}{3} \, \EO \, \mathbb{I}      \label{eq:Ptens0_diffusion}
  \end{empheq}
\end{subequations}
In addition, within the second-order equilibrium diffusion approximation \citep{Mihalas_and_Mihalas_book_1984, Castor_book_2004},
the comoving-frame radiation energy $\EO$ has the following orders of magnitude, in, respectively, the static diffusion regime, and the dynamic diffusion regime:
\begin{subequations} \label{eq:E0_diffusion_regime_static_dynamic}
  \begin{empheq}[left=\text{diffusion\quad}{\empheqlbrace}]{alignat=10}
      &\; \EO & &\overset{\text{static}}{\underset{\text{diffusion}}{\sim}} &\, & \frac{4\, \pi \, \BO}{c} \, \left[ 1 \pm \left( \frac{1}{\tau}\right)^{2} \right]   \label{eq:E0_static_diffusion}  \\
      &\; \EO & &\overset{\text{dynamic}}{\underset{\text{diffusion}}{\sim}} &\, & \frac{4\, \pi \, \BO}{c} \, \left[ 1 \pm \left( \frac{\beta}{\tau}\right) \right]  \label{eq:E0_dynamic_diffusion}
  \end{empheq}
\end{subequations}
Using the results above, we can now estimate the orders of magnitude of each term of Eq.~(\ref{eq:Erad_integrated_approx})
and (\ref{eq:Frad_integrated_approx}).
We globally follow the usual approaches described in \citet{Mihalas_and_Klein_1982,Mihalas_and_Mihalas_book_1984,Stone_et_al_1992,Krumholz_et_al_2007,Skinner_and_Ostriker_2013}.
As did \citet{Krumholz_et_al_2007} and \citet{Skinner_and_Ostriker_2013}, we adopt the characteristic length $\ell$ in the three asymptotic regimes, for the fluid and radiation quantities.
We adopt the characteristic timescale $\tf$ for the fluid quantities.
We adopt the following characteristic timescales for the radiation quantities:
$\tr$ in the free-streaming regime, $\td$ in the static diffusion regime, and $\tf$ in the dynamic diffusion regime.
The spatial derivatives are characterized by $1/\ell$. The time derivatives are characterized by $1/\text{(characteristic timescale)}$.\\

\noindent Table~\ref{tab:Erad_scaling_terms} and Table~\ref{tab:Mrad_scaling_terms} display the relative scaling of each term in, respectively,
Eq.~(\ref{eq:Erad_integrated_approx}) and Eq.~(\ref{eq:Frad_integrated_approx}). In each asymptotic regime, the normalization parameter is indicated in the last column.
We also have sorted, in each regime, the terms according to their relative orders of magnitude, from the largest ones (rank~1), to the
smallest ones (rank~3 or rank~4). The tables show that the terms with acceleration (~(f) in Table~\ref{tab:Erad_scaling_terms},
(g) and (h) in Table~\ref{tab:Mrad_scaling_terms}) are from one to three orders of magnitude
smaller than the leading terms. Then, one can safely always neglect them, as already stated by \citet{Mihalas_and_Mihalas_book_1984}.
Moreover, the terms that account for the time derivative of the radiation momentum (~(c) in Table~\ref{tab:Erad_scaling_terms}),
and of the radiation pressure (~(f) in Table~\ref{tab:Mrad_scaling_terms}), are from one to two orders
of magnitude smaller than the leading terms. Then, one can also safely always neglect them.
In Table~\ref{tab:Erad_scaling_terms}, the relative order of magnitude of the radiation energy source - sink term,
$- \, c \, \GOO$, cannot be estimated a priori in non-LTE (it is $\tau$ in LTE). That is why, its rank is considered to be one by default.
Finally, in Table~\ref{tab:Mrad_scaling_terms}, the terms (c), (d), and (e) are from one to two orders of magnitude smaller than
the leading terms. So, we can neglect them, to get a solution valid to $O(1)$.

\paragraph{The radiation moment equations to $O(1)$}\mbox{}\\ \label{paragraph:O_1_radiation_moment_equations}

The discussion above on the orders of magnitude suggests that, 
if we keep only the terms that are leading in at least one of the three asymptotic regimes,
then the total non-LTE radiation moment equations are:
\begin{subequations}
   \begin{equation} \label{eq:Erad_integrated_approx_O_1}
        \frac{\dr \EO}{\dr t} + \nabla \cdot \FvecO  + \nabla \cdot \left(\EO \, \vvec \right) + \PtensO : \nabla \, \vvec = \, \LO - c \, \kOP   \, \EO \:, \\
   \end{equation}
   \begin{equation} \label{eq:Frad_integrated_approx_O_1}
     \frac{\dr}{\dr t} \left(\frac{\FvecO}{c^{2}} \right) + \nabla \cdot \PtensO \, = \, - \, \frac{1}{c} \, \chiOR  \, \FvecO.
   \end{equation}
\end{subequations}
Note that, if we want to solve the radiation moment equations to the next order, i.e., by keeping the terms that are
$O(\beta)$ relative to the leading terms, then,
we should add term~$(c)$ from Table~\ref{tab:Erad_scaling_terms} in the radiation energy equation,
and terms~$(c)$, $(d)$, $(e)$, and $(f)$ from Table~\ref{tab:Mrad_scaling_terms} in the radiation momentum equation.

\begin{table*}[ht]
\small
\begin{center}
\caption{Relative sizes of terms in the comoving-frame radiation energy equation (\ref{eq:Erad_integrated_approx}) $\left({\rm erg}\;{\rm cm^{-3}}\;{\rm s^{-1}}\right)$} \label{tab:Erad_scaling_terms}
\vspace{0.2in}
\begin{tabular}{llcccccccc}
\hline\hline\\
\multicolumn{2}{l}{\textbf{Radiation energy}}               &  $\frac{\dr \EO}{\dr t}$ & $\nabla \cdot \FvecO$       & $\vvec \cdot  \frac{\dr}{\dr t} \left( \frac{\FvecO}{c^{2}} \right)$ & $\nabla \cdot \left(\EO \, \vvec \right)$ & $\PtensO : \nabla \, \vvec$ & $2 \, \FvecO  \cdot \frac{\avec}{c^{2}}$ & $- \, c \, \GOO$ & Normalization \\[0.2cm]
\multicolumn{2}{l}{\textbf{(tag)}}                          &  (a) & (b) & (c) & (d) & (e) & (f) & (g)  & \\[0.2cm]
\hline
\rule{0pt}{15pt}
   \multirow{2}{*}{\textbf{free streaming}} & $t_{r} \sim \frac{\ell}{c}$  & 1 & 1 & $\beta$ & $\beta$ & $\beta$ & $\beta^{2}$ & $\LO \frac{t_{r}}{\EO} (\underset{{\rm LTE}}{\to}\tau)$ & \multirow{2}{*}{$\times \left(\frac{\EO}{t_{r}}\right)$} \\[0.2cm]
                                                             &\textbf{rank}& 1 & 1 &     2   &    2    &    2    &      3      &         1                                                 \\[0.2cm]
\hline
\rule{0pt}{15pt}
   \textbf{static diffusion} & $t_{d} \sim \frac{\ell^{2}}{c \, \lambdap}$ & 1 & 1 & $\frac{\beta}{\tau}$ & $\beta \tau$ & $\beta \tau$ & $\beta^{2}$ & 1 & \multirow{2}{*}{$\times \left(\frac{\EO}{t_{d}}\right)$} \\[0.2cm]
   $t_{d} \ll t_{f}$                                         &\textbf{rank}& 1 & 1 &           3          &       2      &       2      &     4       & 1   \\[0.2cm]
\hline
\rule{0pt}{15pt}
   \textbf{dynamic diffusion} & $t_{f} \sim \frac{\ell}{\varv}$            & 1 & $\frac{1}{\beta\tau}$ & $\frac{\beta}{\tau}$ & 1 & 1 & $\frac{\beta}{\tau}$ & 1 & \multirow{2}{*}{$\times \left(\frac{\EO}{t_{f}}\right)$} \\[0.2cm]
   $t_{f} \ll t_{d}$                                         &\textbf{rank}& 1 &        2              &           3          & 1 & 1 &        3             & 1   \\[0.2cm]
\hline
\end{tabular}
\bigskip
\end{center}
\end{table*}

\begin{table*}[ht]
\small
\begin{center}
\caption{Relative sizes of terms in the comoving-frame radiation momentum equation (\ref{eq:Frad_integrated_approx}) $\left({\rm dyn}\;{\rm cm^{-3}}\right)$} \label{tab:Mrad_scaling_terms}
\vspace{0.2in}
\begin{tabular}{llcccccccccc}
\hline\hline\\
\multicolumn{2}{l}{\textbf{Radiation momentum}}             &  $\frac{\dr}{\dr t} \left(\frac{\FvecO}{c^{2}}\right)$ & $\nabla \cdot \PtensO$   & $\left(\vvec \cdot \nabla \right) \left(\frac{\FvecO}{c^{2}}\right)$ & $\left(\frac{\FvecO}{c^{2}} \cdot \nabla\right) \vvec$ & $\left(\nabla \cdot \vvec \right) \left(\frac{\FvecO}{c^{2}}\right)$ & $\frac{\dr \PtensO}{\dr t} \frac{\vvec}{c^{2}}$ & $\frac{1}{c^{2}} \left( \PtensO \, \avec \right)$ & $\frac{1}{c^{2}} \left( \EO \, \avec \right)$ & $-\GvecO$ & Normalization \\[0.2cm]
\multicolumn{2}{l}{\textbf{(tag)}}                          & (a) & (b) & (c) & (d) & (e) & (f) & (g) & (h) & (i) & \\[0.2cm]
\hline
\rule{0pt}{15pt}
   \multirow{2}{*}{\textbf{free streaming}} & $t_{r} \sim \frac{\ell}{c}$  & 1 & 1 & $\beta$ & $\beta$ & $\beta$ & $\beta$ & $\beta^{2}$ & $\beta^{2}$ & $\tau$   & \multirow{2}{*}{$\times \left(\frac{\FO}{c \, \ell}\right)$} \\[0.2cm]
                                                             &\textbf{rank}& 1 & 1 & 2 & 2 & 2 & 2 & 3 & 3 & > 1                         \\[0.2cm]
\hline
\rule{0pt}{15pt}
   \textbf{static diffusion} & $t_{d} \sim \frac{\ell^{2}}{c \, \lambdap}$ & $\frac{1}{\tau^{2}}$ & 1 & $\frac{\beta}{\tau}$ & $\frac{\beta}{\tau}$ & $\frac{\beta}{\tau}$ & $\frac{\beta}{\tau}$ & $\beta^{2}$ & $\beta^{2}$ & 1 & \multirow{2}{*}{$\times \left(\frac{\FO}{c \, \lambdap}\right)$} \\[0.2cm]
   $t_{d} \ll t_{f}$                                         &\textbf{rank}& 2 & 1 & 3 & 3 & 3 & 3 & 4 & 4 &  1                         \\[0.2cm]
\hline
\rule{0pt}{15pt}
   \textbf{dynamic diffusion} & $t_{f} \sim \frac{\ell}{\varv}$            & $\frac{\beta}{\tau}$ & 1 & $\frac{\beta}{\tau}$ & $\frac{\beta}{\tau}$ & $\frac{\beta}{\tau}$ & $\beta^{2}$ & $\beta^{2}$ & $\beta^{2}$          & 1 & \multirow{2}{*}{$\times \left(\frac{\FO}{c \, \lambdap}\right)$} \\[0.2cm]
   $t_{f} \ll t_{d}$                                         &\textbf{rank}& 3 & 1 & 3 & 3 & 3 & 2 & 2 & 2 & 1                         \\[0.2cm]
\hline
\end{tabular}
\bigskip
\end{center}
\end{table*}

\subsubsection{The flux-limited diffusion approximation} \label{subsubsec:approx_4_FLD}

To close the system, we apply the flux-limited (FLD) approximation
\citep{Alme_and_Wilson_1973}. The FLD approach, which has been widely used in many RHD codes
(cf. \S~\ref{sec:introduction} for a review) consists in replacing the radiation momentum equation with
a Fick's law of diffusion that links the comoving total radiation flux to the comoving total radiation
energy through a radiation diffusion coefficient $K$, as written below:
\begin{equation}  \label{eq:FLD_Fick_law}
   \FvecO \, = \, - K \, \nabla \EO   \qquad \text{with} \quad K \, \equiv \, \frac{c \, \lambda}{\chiOR}
\end{equation}
where $\lambda$ (not to be confused with the photon mean free path $\lambdap(\nuO)$ defined in \S~\ref{paragraph:the_3_asymptotic_regimes})
is the so-called flux-limiter, a dimensionless quantity that should be defined so that the relation between the radiation flux and the radiation energy
is correct in the optically thin and optically thick asymptotic regimes. In other words, we must choose $\lambda$ so that:
\begin{equation} \label{eq:lambda_limits}
     \left\{ 
         \begin{alignedat}{2}
                &\lambda \to & \, &1/R  \quad \text{in an optically thin medium} \\
                &\lambda \to & \, &1/3  \quad \text{in an optically thick medium}
         \end{alignedat}
         \right.
\end{equation}
where the dimensionless quantity $R$ is defined as follows:
\begin{equation} \label{eq:R_definition}
     R \, \equiv \, \frac{\left| \nabla \EO \right|}{\chiOR \, \EO}  \sim \frac{1}{\tau}
     \left\{ 
         \begin{alignedat}{2}
                &\xrightarrow[\text{optically thin}]{}  \; & &\infty\\
                &\xrightarrow[\text{optically thick}]{} \; & &0
         \end{alignedat}
         \right.
\end{equation}
This way, we recover the asymptotic relations between $\EO$ and $\FvecO$, in an optically thin medium, Eq.~(\ref{eq:Fvec0_streaming}), and
in an optically thick medium, Eq.~(\ref{eq:Fvec0_equilibrium_diffusion}).
The flux-limiter is then defined as a function of $R$. Different functions are proposed in the literature.
The following three ones, respectively suggested by \citet{Levermore_and_Pomraning_1981} (Eq.~\ref{eq:lambda_Levermore_Pomraning_1981}),
\citet{Minerbo_1978} (Eq.~\ref{eq:lambda_Minerbo_1978}), and \citet{Kley_1989} (Eq.~\ref{eq:lambda_Kley_1989}),
were implemented in PLUTO by \citet{Kolb_et_al_2013}:
\begin{subequations}
   \begin{equation} \label{eq:lambda_Levermore_Pomraning_1981}
      \lambda(R) \, = \, \frac{1}{R}\left(\coth R - \frac{1}{R}\right) \\
   \end{equation}
   \begin{equation} \label{eq:lambda_Minerbo_1978}
     \lambda(R) \, = \,
       \left\{ 
         \begin{alignedat}{2}
                &\frac{2}{3 + \sqrt{9+12 R^2}} & \quad &0 \leq R \leq \frac{3}{2}\\
                &\frac{1}{1+R+\sqrt{1+2R}}     & \quad &\frac{3}{2} < R < \infty\\
         \end{alignedat}
         \right.
   \end{equation}
   \begin{equation} \label{eq:lambda_Kley_1989}
       \lambda(R) \, = \,
         \left\{ 
            \begin{alignedat}{2}
		&\frac{2}{3 + \sqrt{9+10 R^2}} &          \quad &0 \leq R \leq 2\\
		&\frac{10}{10 R+ 9 + \sqrt{180 R + 81}} & \quad &2 < R < \infty\\
         \end{alignedat}
         \right.
   \end{equation}
\end{subequations}

\noindent Finally, one needs a closure relation to relate the radiation pressure to the radiation energy.
The most commonly used one (e.g., by \citealt{Turner_and_Stone_2001}, \citealt{Krumholz_et_al_2007}, \citealt{Zhang_et_al_2011},
\citealt{Zhang_et_al_2013}) is that provided by \citet{Levermore_1984}:
 \begin{equation} \label{eq:P0_vs_E0}
     \PtensO \, = \, \frac{\EO}{2} \, \left[\left(1-f\right) \mathbb{I} + \left(3 f - 1\right) \nvecO \otimes \nvecO \right]
 \end{equation}
where $f$ is the Eddington factor, a dimensionless quantity, related to $\lambda$ and $R$ as follows:
 \begin{equation} \label{eq:f_lambda_R}
     f \, = \, \lambda + \lambda^{2} R^{2}
     \quad
     \left\{ 
         \begin{alignedat}{2}
                &\xrightarrow[\text{optically thin}]{}  \; & &1 \\
                &\xrightarrow[\text{optically thick}]{} \; & &\frac{1}{3}
         \end{alignedat}
         \right.
 \end{equation}
where the limits are inferred from (\ref{eq:lambda_limits}) and (\ref{eq:R_definition}).
Then, we verify that (\ref{eq:P0_vs_E0}) recovers the following asymptotic relations: (\ref{eq:Ptens0_streaming})
in an optically thin medium, and (\ref{eq:Ptens0_diffusion}) in an optically thick medium.

\subsection{The non-LTE RMHD equations solved by PLUTO} \label{subsec:PLUTO_RMHD_equations}

We have implemented in a module, which is coupled to PLUTO, the radiation terms that account for a non-LTE radiation hydrodynamics (RHD) description of a flow,
by expanding the LTE RHD equations \citep{Kolb_et_al_2013} to the more general non-LTE regime.
Since PLUTO is a MHD code, we have actually enhanced its capabilities, so that it is now a 3D non-LTE radiation magnetohydrodynamics (RMHD) code.

We use the RHD equations within the approximations detailed throughout \S~\ref{subsec:NLTE_approx_RHD}.
The full set of 3D nonrelativistic RMHD equations solved by PLUTO is based on the full MHD equations \citep{Mignone_et_al_2007, Mignone_et_al_2012},
and on Equations~(\ref{eq:continuity_integrated_approx}), (\ref{eq:matter_momentum_integrated_approx}), (\ref{eq:matter_energy_integrated_approx}),
(\ref{eq:Erad_integrated_approx_O_1}), and (\ref{eq:FLD_Fick_law}). The system is written as

\begin{subequations}
   \begin{equation} \label{eq:continuity_PLUTO}
      \left\{
      \begin{alignedat}{2}
   &\frac{\dr \rho}{\dr t} + \nabla \cdot \left( \rho \, \vvec\right) = 0 \:,\\
    \end{alignedat}
      \right.
   \end{equation}
   \begin{equation} \label{eq:matter_momentum_PLUTO}
      \left\{
      \begin{alignedat}{2}
   &\frac{\dr}{\dr t}\left( \rho \, \vvec \right) +  \nabla \cdot \left[ \rho \, \vvec \otimes \vvec - \Bvec \otimes \Bvec + \left(p + \tfrac{1}{2}\,B^{2} \right) \, \mathbb{I} \right] \\
   & = \rho \, \gvec + \nabla \cdot \visctens + \frac{1}{c} \, \chiOR  \, \FvecO  \:, \\
    \end{alignedat}
      \right.
   \end{equation}
   \begin{equation} \label{eq:matter_energy_PLUTO}
      \left\{
      \begin{alignedat}{2}
   &\frac{\dr}{\dr t} \left( \Em + \tfrac{1}{2} \, \rho \, \varv^2 + \tfrac{1}{2}\,B^{2}\right) \\
   &+ \nabla \cdot \left[ \left( \Em + \tfrac{1}{2} \, \rho \, \varv^2 + \tfrac{1}{2}\,B^{2} + p + \tfrac{1}{2}\,B^{2}\right) \, \vvec - \left(\vvec \cdot \Bvec\right) \Bvec \right] \\
   & = \rho \, \gvec \cdot \vvec  + \nabla \cdot \left[ -\Fcond + \visctens \, \vvec - \left( \eta \, \Jvec \right) \times \Bvec \right] \\
   & \hspace{0.5cm} + \frac{1}{c} \, \chiOR  \, \FvecO \cdot \vvec + c \, \kOP   \, \EO - \LO \:, 
    \end{alignedat}
      \right.
   \end{equation}
   \begin{equation} \label{eq:Erad_integrated_approx_O_1_PLUTO}
      \left\{
      \begin{alignedat}{2}
        \frac{\dr \EO}{\dr t} + \nabla \cdot \FvecO  = \, \LO - c \, \kOP   \, \EO \:, \\
    \end{alignedat}
      \right.
   \end{equation}
   \begin{equation} \label{eq:FLD_Fick_law_PLUTO}
      \left\{
      \begin{alignedat}{2}
        \FvecO \, = \, - \left(\frac{c \, \lambda}{\chiOR}\right) \, \nabla \EO \:, \\
    \end{alignedat}
      \right.
   \end{equation}
   \begin{equation} \label{eq:Faraday_law}
      \left\{
      \begin{alignedat}{2}
        \frac{\dr \Bvec}{\dr t} + \nabla \cdot \left(\vvec \otimes \Bvec - \Bvec \otimes \vvec \right) \, = \, - \nabla \times \left( \eta \, \Jvec \right)  \:, \\
    \end{alignedat}
      \right.
   \end{equation}
   \begin{equation} \label{eq:EOS}
      \left\{
      \begin{alignedat}{2}
        p = \rho \, \frac{\kB \, T}{\mu \, \mH} \:, \quad \text{(EOS)}\\
    \end{alignedat}
      \right.
   \end{equation}
   \begin{equation} \label{eq:gas_energy_temperature}
      \left\{
      \begin{alignedat}{2}
        \Em = \frac{p}{\gamma -1} = \rho \, \frac{\kB}{\left(\gamma - 1\right) \, \mu \, \mH} \, T\:. \quad \text{(caloric EOS)}
    \end{alignedat}
      \right.
   \end{equation}
\end{subequations}
where $B$ is the magnetic field, $\lambda$ is the flux-limiter (defined in \S~\ref{subsubsec:approx_4_FLD}), $\kB$ is the Planck constant,
$\gamma$ is the ratio of specific heats, $\mu$ is the mean molecular weight, $\mH$ is the standard mass of one atom of hydrogen,
$\Fcond$ is the conductive flux, $\visctens$ is the viscous tensor, $\eta$ is the magnetic resistivity tensor, and $\Jvec \equiv \nabla \times \Bvec$
is the current density (for more information on these quantities, see \citealt{Mignone_et_al_2007, Mignone_et_al_2012,Orlando_et_al_2008}).

The opacities $\kOP$, $\chiOR$, and the radiative losses $\LO$ can be provided separately.
In our case, we use databases pre-calculated as functions of density and temperature in a non-LTE regime (cf. \S~\ref{subsubsec:approx_mean_opacities}).

Note that an additional approximation is made in the radiation energy equation (\ref{eq:Erad_integrated_approx_O_1_PLUTO}) solved by PLUTO.
The terms $\nabla \cdot \left(\EO \, \vvec \right)$ and $\PtensO : \nabla$ are missing, compared to (\ref{eq:Erad_integrated_approx_O_1}).
They were already missing in the LTE RHD version by \citet{Kolb_et_al_2013} (cf. their equation 4). Table~\ref{tab:Erad_scaling_terms} reveals that these
two terms are leading terms only in the dynamic diffusion regime, but are of second order in the static diffusion regime and in the free streaming regime.
So, at this stage, our module cannot be used in the dynamic diffusion regime.
We plan to implement these two terms in a future upgraded version of our RHD module.

\section{Non-LTE opacities and radiative power loss: theoretical model} \label{sec:opacity}

Our module needs, at a given set of density and temperature $(\rho,T)$,
and for a given composition of the plasma, the following input data: the radiation emission $\LO$,
the Planck mean opacity $\kOP$, and the Rosseland mean opacity $\chiOR$.

Databases have been generated by \cite{Rodriguez_et_al_2018} in a non-LTE regime, but also in LTE regime.
Our module uses three tables, ($\LO, \koP, \koR$) as functions of ($\rho$, $T$),
for a plasma with solar-like abundances, where $\koP$ and $\koR$ are (${\rm cm^{2}}\;{\rm g^{-1}}$):
\begin{subequations}
  \begin{empheq}[left={\empheqlbrace}]{alignat=4}
      &\koP  \, & &\equiv \, & &\frac{\kOP}{\rho}  \label{eq:koP_def} \\
      &\koR  \, & &\equiv \, & &\frac{\chiOR}{\rho}  \label{eq:koR_def}
  \end{empheq}
\end{subequations}

In this section we briefly summarize the features of the theoretical model, whose details are explained in \cite{Rodriguez_et_al_2018}.
Note that our module can read any other set of data ($\LO, \koP, \koR$) that would be provided by the user.

Plasma radiative properties depend on the plasma level populations and atomic properties.
In the present work, the atomic quantities of the different chemical elements of the multi-component plasma,
such as the relativistic energy levels, wave functions, oscillator strengths and photoionization cross sections,
were calculated using the FAC code \citep{Gu_et_al_2008}, in which a fully relativistic approach based on the Dirac equation is solved.
The atomic calculations were carried out in the relativistic detailed configuration account \citep{Bauche_et_al_1987}.
The atomic configurations selected for each ion in the plasma mixture were those with energies within two times the ionization energy
of the ground configuration of the ion.

The atomic level populations were obtained assuming the plasma in steady-state.
This approach is valid when the characteristic time of the must relevant atomic process in the plasma
is considerably shorter than the time associated to changes in the plasma density and temperature, i.e.,
the characteristic time of the plasma evolution. When this criterion is fulfilled, the atomic processes are fast enough
to distribute the atomic level populations in the plasma before the density and temperature of the plasma change.
This approach is commonly used to obtain the atomic level populations in the plasma needed in the calculation of
radiative properties databases for radiation-hydrodynamics simulations. In the steady-state approximation,
the population density of the atomic level $i$ of the ion with charge state $\zeta$, denoted as $N_{\zeta i}$ is obtained
by solving the set of rate equations implemented in a collisional-radiative steady-state (CRSS) model, given by
\begin{equation}
    \sum_{\zeta' j} N_{\zeta' j}(\vec{r},t)R^+_{\zeta'j \rightarrow \zeta i} - \sum_{\zeta' j} N_{\zeta i}(\vec{r},t)R^-_{\zeta i \rightarrow \zeta' j} = 0
    \label{level}
\end{equation}
Two complementary equations have to be satisfied together with Eq.~(\ref{level}). First, the requirement that the sum of
all the partial densities equals the total ion density and, second, the charge neutrality condition in the plasma.
In Eq.~(\ref{level}), $R^+_{\zeta' j \rightarrow \zeta i}$ and $R^-_{\zeta' j \rightarrow \zeta i}$ take into account all
the atomic processes which contribute to populate and depopulate the atomic configuration $\zeta_i$, respectively.
The atomic processes included in the CRSS model were collisional ionization \citep{Lotz_1968} and three-body recombination,
spontaneous decay \citep{Gu_et_al_2008}, collisional excitation \citep{van_Regemorter_1962} and deexcitation,
radiative recombination \citep{Kramers_1923}, autoionization and electron capture \citep{Griem_1997_Book}.
The rates of the inverse processes were obtained through the detailed balance principle. In the simulations carried out
in this work, the plasma was assumed as optically thin. The effect of the plasma environment on the population of the
atomic levels was modeled through the depression of the ionization potential or continuum lowering, which can reduce
the number of bound states available. The formulation developed by \cite{Stewart_et_al_1966} was applied.
In the present CRSS model, the ions were considered to be at rest. On the other hand, in the calculation of the rates
of the atomic processes, a Maxwell-Boltzmann distribution for the free electrons was assumed.
For electron densities between $10^{11}$ and $10^{14}$ cm$^{-3}$ and electron temperatures lower than $200$ eV,
the electron mean free paths range between $3.33\times10^5$ and $25.8$~cm \citep{Rodriguez_et_al_2018}.
This property provides an estimation of the average volume needed for the free electrons to thermalize.
For the range of plasma conditions analyzed in this work, the Fermi-Dirac distribution is not necessary.
The CRSS model described is implemented in MIXKIP code \citep{Espinosa_et_al_2017}.

As said before, in this work a multi-component plasma was simulated. The chemical elements considered in the mixture were
H, He, C, N, O, Ne, Na, Mg, Al, Si, S, Ar, Ca and Fe,
and the solar-like abundances provided by \cite{Asplund_et_al_2009} were used. For a given electron density and temperature,
the CRSS model is solved for each element without considering any atomic processes that connect different chemical elements.
However, they were coupled through the common pool of free electrons since the plasma level populations
of each chemical element have to be consistent with the common electron density. This fact ensures the plasma neutrality \citep{Klapisch_2013}.

\begin{figure*}
    \centering
    \includegraphics[scale = 0.8]{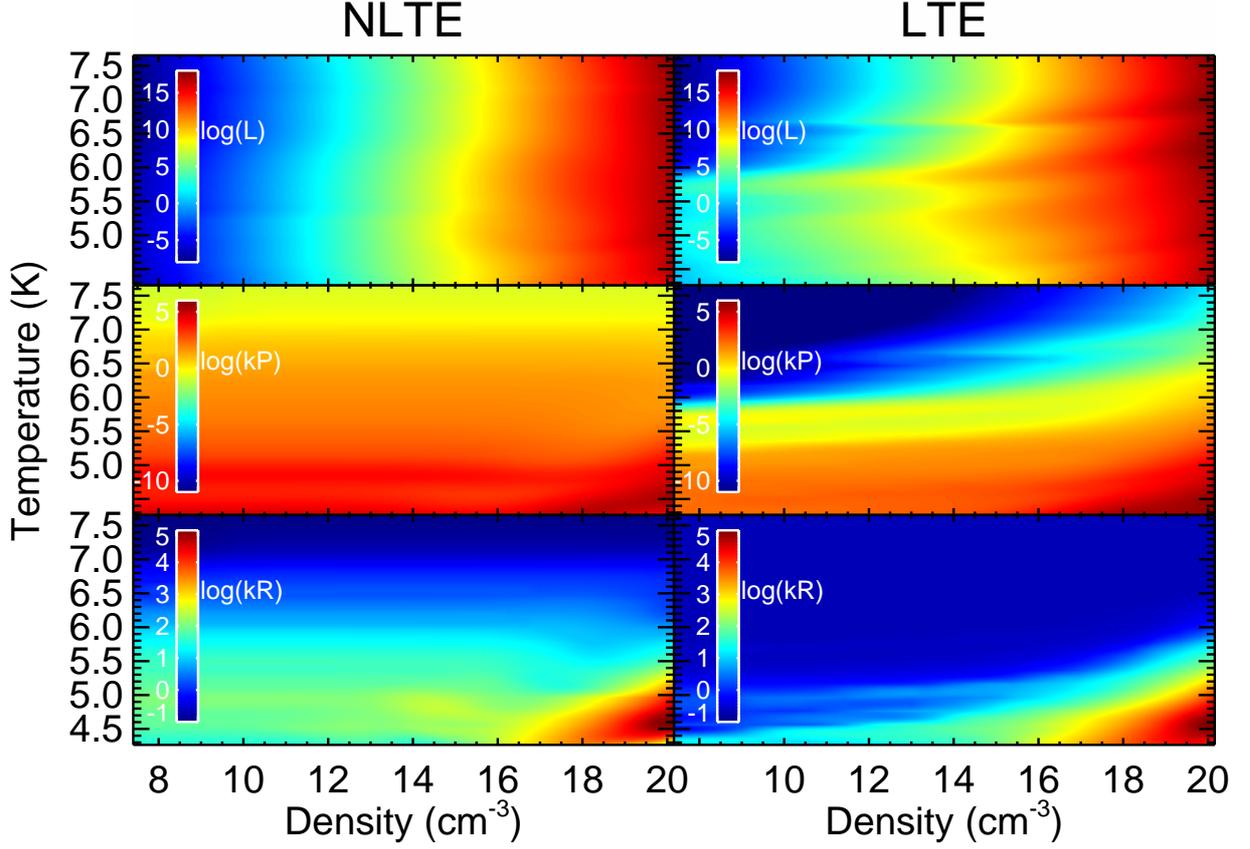}
    \caption{Total radiation emission (${\rm erg}\;{\rm cm^{-3}}\;{\rm s^{-1}}$) (top panels),
             Planck opacities (${\rm cm^{2}}\;{\rm g^{-1}}$) (mid panels),
      and Rosseland opacities (${\rm cm^{2}}\;{\rm g^{-1}}$) (bottom panels),
             in a non-LTE regime (left), and in LTE regime (right). See Section~\ref{sec:opacity} for more details.}
    \label{fig:opacities_NLTE_vs_LTE}
\end{figure*}

Once the atomic data and level populations were obtained, the monochromatic absorption coefficient and emissivity
of each chemical element $m$, $\kappa_m(\nu)$ and $j_m(\nu)$, respectively, with $\nu$ the photon frequency,
were calculated using the RAPCAL code \citep{Rodriguez_et_al_2008,Rodriguez_et_al_2010}.
Both coefficients include the bound-bound, bound-free and free-free contributions.
The radiative transition rates were calculated in the electric dipole approximation using the FAC code.
The oscillator strengths included a correction to take into account configuration interaction effects due to
the mix between relativistic configurations that belong to the same non-relativistic one.
The photoionization cross sections were calculated in the distorted wave approach. Complete redistribution hypothesis
was assumed for the line profile which included natural, Doppler, electron-impact \citep{Dimitrijevic_et_al_1987} and
UTA broadenings \citep{Bauche_et_al_1987}. The line-shape function was applied with the Voigt profile that incorporated
all these broadenings. For the free-free contributions, the Kramers semi-classical expression for the inverse bremsstrahlung
cross section was used \citep{Rose_1992}. In order to determine the opacity, $k(\nu)$, from the absorption coefficient,
the scattering of photons was also taken into account, and, in RAPCAL, this was approximated using the Thomson scattering cross section \citep{Rutten_1995}.
The monochromatic opacities and emissivities of the mixture were obtained from the weighted contributions of the different chemical elements
\begin{subequations}
   \begin{alignat}{2}
    & k(\nu) = \frac{1}{\rho}\sum_m X_m \, \kappa_m(\nu) \\
    & j(\nu) = \sum_m X_m \, j_m(\nu)
   \end{alignat}
\end{subequations}
where $\rho$ is the mass density of the mixture. From the monochromatic opacities of the mixture,
the Planck $\koP$ and Rosseland $\koR$ mean opacities were calculated, as well as the radiative power loss, or radiation emission, $\LO$,
from the monochromatic emissivity.
Figure~\ref{fig:opacities_NLTE_vs_LTE} portrays maps of $\LO$, $\koP$, and $\koR$, vs. free electron density $({\rm cm^{-3}})$
and temperature $({\rm K})$ in logarithmic scale, in a non-LTE regime, and in LTE regime.

Both MIXKIP and RAPCAL codes have been successfully tested with experimental results and numerical simulations for plasmas of
single elements included in the mixture analyzed in this work \citep{Espinosa_et_al_2017,Rodriguez_et_al_2008,Rodriguez_et_al_2010},
both in LTE and non-LTE, and for the plasma mixture in LTE simulations \citep{Rodriguez_et_al_2018}.
A more detailed explanation of both codes can be found in \citet{Rodriguez_et_al_2018}.

\section{Implementation of the radiation terms} \label{sec:implementation}

We explain, here, how the radiative terms of the physical equations are implemented in the code.
We have followed and upgraded to the non-LTE case the techniques described for the LTE regime in \cite{Kolb_et_al_2013}
 (see also \citealt{Commercon_et_al_2011a}, \citealt{van_der_Holst_et_al_2011}, \citealt{Zhang_et_al_2011}, and \citealt{Zhang_et_al_2013}).
We use an operator-split method. As detailed in this section, advancing the variables during a time step is made in two substeps,
an explicit one (\S~\ref{subsec:substep_1}), and an implicit one (\S~\ref{subsec:substep_2}). The latter involves the radiation emission $\LO$, which
is an analytical function of temperature in LTE ($\LO = \koP \, \rho\, c\, a_{R} \, T^{4}$: cf. \S~\ref{subsubsec:approx_mean_opacities}),
but has no analytical expression in non-LTE, and is stored in databases (cf. \S~\ref{sec:opacity}).
Our original contribution, with respect to the usual LTE implementations, consists in
handling this non-analytical radiation emission term in the treatment of the implicit scheme, as explained below in \S~\ref{subsec:substep_2}. 

\subsection{Reformulation of the equations} \label{subsec:reformulation}

For the sake of simplicity, we remove from the full system solved by PLUTO, (\ref{eq:continuity_PLUTO}) - (\ref{eq:gas_energy_temperature}),
the terms depending on the magnetic field, and rewrite it in a simplified version, for an inviscid fluid without heat-conduction, but subject to external gravity
(even though the following discussion can be applied to the full RMHD system):
\begin{subequations}
   \begin{alignat}{2}
      &\frac{\dr \rho}{\dr t} + \nabla \cdot \left( \rho \, \vvec\right) = 0 \label{eq:Mass_c} \\
      &\frac{\dr}{\dr t}\left( \rho \, \vvec \right) +  \nabla \cdot \left( \rho \, \vvec \otimes \vvec + p \, \mathbb{I} \right) = \rho \, \gvec + \frac{\rho \, k_R}{c} \, \Fvec \label{eq:Momentum_c} \\ 
      &\frac{\dr}{\dr t} \left( \Em + \tfrac{1}{2} \, \rho \, \varv^2 \right) +
          \nabla \cdot \left[ \left( \Em + \tfrac{1}{2} \, \rho \, \varv^2 + p \right) \, \vvec \right] = \label{eq:Plasma_energy_c} \\
      &   \hspace{3cm}  \rho \, \gvec \cdot \vvec + \frac{\rho \, k_R}{c} \, \Fvec \cdot \vvec + k_P \, \rho \, c \, E - L \nonumber \\ 
      &\frac{\partial E}{\partial t} + \nabla \cdot \Fvec = L - k_P \, \rho \, c \, E \label{eq:Er_conservation} \\
      &\Fvec = - \frac{c \, \lambda}{k_R \, \rho} \nabla E \label{eq:FLD_relation} \\
      &p = \rho \, \frac{k_B \, T}{\mu \, \mH} \label{eq:EOS_c}\\
      &\Em = \frac{p}{\gamma -1} = \rho \, \frac{\kB}{\left(\gamma - 1\right) \, \mu \, \mH} \, T \label{eq:gas_energy_temperature_c}
   \end{alignat}
\end{subequations}
where we have omitted, for economy of notation, the subscript $0$ in the radiation quantities.

The above system contains 11 unknowns: six principal variables $\left(\rho,\vvec, p, E\right)$ that
are determined by solving the six equations (\ref{eq:Mass_c}), (\ref{eq:Momentum_c}), (\ref{eq:Plasma_energy_c}), and (\ref{eq:Er_conservation}),
and five variables $\left(T,\Em,\Fvec\right)$ that can be inferred from the six principal ones from the FLD relation (\ref{eq:FLD_relation}),
the EOS (\ref{eq:EOS_c}), and the caloric EOS (\ref{eq:gas_energy_temperature_c}).

At each time step $\dtn = t^{n+1} - t^{n}$, determined by the CFL condition \citep{Mignone_et_al_2007},
we advance the gas and radiation variables from $t^{n}$ to $t^{n+1}$, by solving the above system
in two consecutive substeps, as described below. We denote as $\left(\rhon,\vn,\pn,\En\right)$ and $\left(\Tn,\Emn,\Fn\right)$ the values of the variables at time $t^{n}$,
and $\left(\rhonpI,\vnpI,\pnpI,\EnpI\right)$ and $\left(\TnpI,\EmnpI,\FnpI\right)$ their values at time $t^{n+1}$, \textit{after} the \textit{two} substeps are completed.

\subsection{Substep 1: explicit hydrodynamics} \label{subsec:substep_1}

In this first substep, PLUTO solves the hyperbolic subsystem made of Equations (\ref{eq:Mass_c}), (\ref{eq:Momentum_c}), (\ref{eq:Plasma_energy_c}),
but without the radiation source (also known as source-sink) terms $k_P \, \rho \, c \, E - L$ in the gas energy equation (\ref{eq:Plasma_energy_c}).
The system to be solved is:
\begin{subequations}
  \begin{empheq}[left=\text{substep 1\quad}{\empheqlbrace}]{alignat=4}
      &\hspace{0.05cm} \frac{\dr \rho}{\dr t} + \nabla \cdot \left( \rho \, \vvec\right) = 0  \label{eq:continuity_subsystem}\\
      &\hspace{0.05cm} \frac{\dr}{\dr t}\left( \rho \, \vvec \right) +  \nabla \cdot \left( \rho \, \vvec \otimes \vvec + p \, \mathbb{I} \right) = \rho \, \gvec - \lambda \, \nabla E \label{eq:gas_momentum_subsystem}\\
      &\hspace{0.05cm} \frac{\dr}{\dr t} \left( \frac{p}{\gamma - 1} + \tfrac{1}{2} \, \rho \, \varv^2 \right) +  \nonumber \\
      &\hspace{0.1cm}  \nabla \cdot \left[ \left( \frac{p}{\gamma - 1} + \tfrac{1}{2} \, \rho \, \varv^2 + p \right) \, \vvec \right] = \label{eq:gas_energy_subsystem}\\
      & \hspace{3.5cm} \rho \, \gvec \cdot \vvec - \lambda \, \vvec \cdot \nabla E  \nonumber 
  \end{empheq}
\end{subequations}
where we have introduced (\ref{eq:FLD_relation}), (\ref{eq:gas_energy_temperature_c}) in (\ref{eq:Momentum_c}), (\ref{eq:Plasma_energy_c}).
A Godunov type algorithm is applied, with several possibilities of Riemann solvers (see \citealt{Mignone_et_al_2007} for details).

We use a $\ast$ superscript to denote the quantities obtained at the completion of this substep. Some of these quantities have intermediate results
that will be updated in the next substep (\S~\ref{subsec:substep_2}).
Starting from $\left(\rhon,\vn,\pn\right)$, PLUTO determines $\left(\rhostar,\vstar,\pstar\right)$, the terms involving the radiation energy density $\En$ being
in the source part of the solved system, and, therefore, not updated.
The corresponding intermediate temperature is obtained from the EOS (\ref{eq:EOS_c}):
\begin{equation} \label{eq:T_intermediate}
   \Tstar \, = \, \frac{\pstar}{\rhostar \, \frac{k_B}{\mu \, \mH}}
\end{equation}

\subsection{Substep 2: implicit radiation diffusion and source terms} \label{subsec:substep_2}

In this second substep, we determine, at time $t^{n+1}$, the radiation energy density $\EnpI$, and the gas temperature $\TnpI$,
by solving the radiation energy equation (\ref{eq:Er_conservation}), and the gas energy equation (\ref{eq:Plasma_energy_c})
without any velocity term (we couple the gas internal energy density evolution rate with the radiation source terms).
The system to be solved in non-LTE is:
\begin{subequations}
  \begin{empheq}[left=\text{substep 2\quad}{\empheqlbrace}]{alignat=4}
      &\frac{\partial E}{\partial t} -\nabla \cdot \left( \frac{c\lambda}{\rho \, k_R}\nabla E\right) \, & &= \, & & L-k_P\,\rho \, c \, E     \label{eq:system_code_a}\\
      &\frac{\partial \Em}{\partial t}                                                                \, & &= \, & & k_P \, \rho \, c \, E - L \label{eq:system_code_b}
  \end{empheq}
\end{subequations}
where the gas energy density $\Em$ is directly related to the gas temperature through the caloric equation of state (\ref{eq:gas_energy_temperature_c}).

Since timescales for radiation are much smaller than timescales for hydrodynamics,
time steps that would be inferred from the CFL condition would lead, if applied to radiation, to impracticable computations. 
Consequently, we use an implicit scheme for solving the system (\ref{eq:system_code_a}) - (\ref{eq:system_code_b}),
with the time step $\dtn$ provided by the explicit hydrodynamics substep~1.

Starting from $\left(\En,\rhostar, \Tstar\right)$, we obtain, at the completion of this substep, $\left(\EnpI,\TnpI\right)$.
The time discretization of (\ref{eq:system_code_a}) and (\ref{eq:system_code_b}) is made according to the following implicit scheme:
\begin{subequations}
  \begin{empheq}[left={\empheqlbrace}]{alignat=4}
      &\frac{\EnpI - \En}{\dtn} - \nabla \cdot \left( \Kstar \, \nabla \EnpI \right) \, & &= \, & & \LnpI - \kPstar \, \rhostar c \, \EnpI  \label{eq:rad_en_time_discretization}\\
      &\rhostar \, \cV \, \frac{\TnpI-\Tstar}{\dtn}                                  \, & &= \, & & \kPstar \, \rhostar c \, \EnpI - \LnpI  \label{eq:gas_en_time_discretization}
  \end{empheq}
\end{subequations}
where, while relating $\Em$ to $\rho$ and $T$ with (\ref{eq:gas_energy_temperature_c}), we assume
that the mass density $\rho$ is not modified by this implicit substep
(following \citealt{Commercon_et_al_2011a}, \citealt{Zhang_et_al_2011}, \citealt{Zhang_et_al_2013}, and \citealt{Kolb_et_al_2013}), and where
\begin{subequations}
  \begin{empheq}[left={\empheqlbrace}]{alignat=4}
      &\cV     \, & &= \, \frac{\kB}{\left(\gamma - 1\right) \, \mu \, \mH}    \label{eq:cv_def}     \\
      &\LnpI   \, & &= \, L\left(\rhonpI,\TnpI\right)                          \label{eq:LnpI_def}   \\
      &\kPstar \, & &= \, k_P\left(\rhostar,\Tstar\right)                      \label{eq:kPstar_def} \\
      &\kRstar \, & &= \, k_R\left(\rhostar,\Tstar\right)                      \label{eq:kRstar_def} \\
      &\Kstar  \, & &= \, \frac{c \, \lambda(\Rstar)}{\kRstar \, \rhostar} \quad \text{with} \quad \Rstar \, = \, \frac{\left| \nabla \En \right|}{\rhostar\, \kRstar \, \En} \label{eq:K_star_def}
  \end{empheq}
\end{subequations}
The radiation diffusion coefficient $K$, the flux-limiter $\lambda$, and the dimensionless quantity $R$ have been defined in \S~\ref{subsubsec:approx_4_FLD}.

The space discretization of (\ref{eq:rad_en_time_discretization}) is obtained from a finite volume method
(now adapted to non-LTE regimes):
\begin{subequations}
   \begin{alignat}{2}
      &\frac{E_{ijk}^{n+1} - E_{ijk}^n}{\dtn} - h_{ijk}(\EnpI,\Kstar) \, = \, L^{n+1}_{ijk}-\kPstarijk \, \rhostar c \, \EnpI_{ijk}  \label{eq:discr_radiation_energy} \\
      &\text{where} \nonumber \\
      &\hijk(\EnpI,\Kstar) \, \equiv \, \nabla \cdot \left( \Kstar \, \nabla \EnpI \right)   \label{eq:discr_rad_diffusion}
   \end{alignat}
\end{subequations}
represents the discretized radiation diffusion term, a linear function of $\EnpI$ given by Equation~(14) of \citet{Kolb_et_al_2013},
and where the indices $i,j,k$ identify the positions of the cell centers in the 3D computational grid.

The radiation emission at time $t^{n+1}$, $\LnpI$ (Eq.~\ref{eq:LnpI_def}), is determined from a $1^{\rm st}$ order Taylor expansion,
starting from the state $(\rhostar,\Tstar)$ obtained from substep~1:
\begin{subequations} 
  \begin{empheq}[left={\empheqlbrace}]{alignat=4}
      &\LnpI = \Lstar + \left(\frac{\partial L}{\partial T}\right)^{\ast}\left(\TnpI-\Tstar\right) + \bcancel{\left(\frac{\partial L}{\partial \rho}\right)^{\ast}\left(\rhonpI-\rhostar\right)} \label{eq:taylor} \\
      &\text{with} \nonumber \\
      &\Lstar   \, = \, L\left(\rhostar,\Tstar\right)                          \label{eq:Lstar_def}   \\
      &\left(\frac{\partial L}{\partial T}\right)^{\ast} \, = \, \left(\frac{\partial L}{\partial T}\right) \left(\rhostar,\Tstar\right);
        \quad \left(\frac{\partial L}{\partial \rho}\right)^{\ast} \, = \, \left(\frac{\partial L}{\partial \rho}\right) \left(\rhostar,\Tstar\right)
  \end{empheq}
\end{subequations}
where we have struck out the term in mass density.
This is because we have assumed that the mass density $\rho$ is not modified by this substep~2.
Moreover, the dependence of the radiation emission $L$ on $T$ is much higher
than that on $\rho$ (see \citealt{Rodriguez_et_al_2018} for more details).

Using (\ref{eq:taylor}) inside (\ref{eq:gas_en_time_discretization}) and (\ref{eq:discr_radiation_energy}),
we obtain, after some algebra (we removed the subscript $i,j,k$ from each term for simplicity): 
\begin{subequations} 
  \begin{empheq}[left={\empheqlbrace}]{alignat=4}
      &\displaystyle \frac{\EnpI-\En}{\dtn} = h(\EnpI,\Kstar) + \frac{\Lstar-\kPstar \, \rhostar c \, \EnpI}{1+\left(\frac{\dtn}{\rhostar \, c_v}\right)\left(\frac{\partial L}{\partial T}\right)^{\ast}}
      \label{eq:final_rad_energy} \\
      &T^{n+1} = T^{n} + \frac{\left(\frac{\dtn}{\rhostar \, c_v}\right)}{1+\left(\frac{\dtn}{\rhostar \, c_v}\right)\left(\frac{\partial L}{\partial T}\right)^\ast}
                         \left( \kPstar \, \rhostar \, c \, \EnpI - \Lstar \right) \label{eq:final_temperature}
  \end{empheq}
\end{subequations}
Eq.~(\ref{eq:final_rad_energy}) represents a linear system whose unknown is the radiation energy density $\EnpI_{ijk}$,
at time $t^{n+1}$, and at all positions $(i,j,k)$ in the computational domain.
The system is solved using the PETSc solver, which is already implemented in the original version of the module;
also, the boundary conditions for the radiation energy density can be periodic, symmetric, reflective, or with fixed values.

Note that such a linearization procedure of a radiative loss function, coupled to an implicit scheme, was previously implemented
by \citet{Cunningham_et_al_2011} (cf. their equations 7 and 8, to be respectively compared to Eq.~(\ref{eq:system_code_b}) and (\ref{eq:system_code_a}) above).
In their application, the formation of star clusters, they use a tabulated function that represents line emission \citep{Cunningham_et_al_2006}
superimposed on the LTE opacity due to dust  \citep{Semenov_2003}.
The detailed system of equations, implemented in the ORION code \citep{Krumholz_et_al_2007}, is presented in \citet{Krumholz_et_al_2011,Krumholz_et_al_2012}.

Once $\EnpI$ is determined, the calculation of the temperature $\TnpI$ is straightforward by simply applying formula (\ref{eq:final_temperature}).
Then, using the approximation $\rhonpI = \rhostar$, we update the Rosseland opacity $\kRnpI$, 
and then the flux-limiter $\lambda(\RnpI)$, therefore the radiation contribution to the right-hand side of Eq.~(\ref{eq:gas_momentum_subsystem}) and
(\ref{eq:gas_energy_subsystem}).
We also immediately obtain the pressure $\pnpI$ from the EOS (\ref{eq:EOS_c}):
\begin{equation} \label{eq:p_np1}
   p^{n+1} \, = \,\rhonpI \, \frac{k_B}{\mu \, \mH} \, \TnpI
\end{equation}
The velocity $\vvec$ is not involved in the equations solved by this substep~2. Then, $\vnpI = \vstar$.
Finally, from the above quantities, we can infer the gas energy $\EmnpI$ by applying (\ref{eq:gas_energy_temperature_c}),
and the radiation flux $\Fvec^{n+1}$ by applying (\ref{eq:FLD_relation}).

\section{Tests} \label{sec:test}

To validate our implementation of the non~LTE equations in the radiation module in PLUTO,
we simulated some test cases, and compared our solutions either to analytical, or semi-analytical results, when they exist,
or, when appropriate, to the LTE version of the code.
All the test cases present in the literature assume the LTE regime.
To have a direct comparison with those tests, we always used our non-LTE discretization scheme (cf. \S~\ref{subsec:substep_2}), but
imposed a radiation emission (or radiative losses) $L$ to be equal to the LTE emission ($L = k_P \, \rho\, c\, a_{R} \, T^{4}$).
Such tests are described below in \S~\ref{subsec:radiation_matter_coupling} and \S~\ref{subsec:radiative_shocks}.

The tests are one-dimensional problems. Even though PLUTO works in one, two or three dimensions, the radiation module was developed only in 3D.
For this reason, to model the test cases we used quasi-one-dimensional domains, which are cuboids with a length much longer than the width or the height.
The tests were made only in Cartesian grid. Since, we did not change the geometrical terms in Eq.~(\ref{eq:final_rad_energy}),
we did not need to check the results using different grids.

To compare the results with \cite{Kolb_et_al_2013}, we used the solver based on the PETSc library for all test cases,
with the GMRES iteration scheme and a block-Jacobi (bjacobi) preconditioner.

\subsection{Radiation matter coupling} \label{subsec:radiation_matter_coupling}
Our objective is to test the correctness of the implementation of Equation~(\ref{eq:system_code_b}), which couples the evolution of the matter (or gas) internal energy $\Em$
with the radiation source terms $k_P \, \rho \, c \, E - L$. This equation is solved with an implicit method, in substep~2 of the operator-split scheme (cf. \S~\ref{subsec:substep_2}).
In particular, this test will enable us to check our linearization procedure of the radiation emission $L$ (Eq.~\ref{eq:taylor}), wich results in the expression of the temperature
$\TnpI$ versus $\Tn$ (Eq.~\ref{eq:final_temperature}). Such a test was first proposed by \cite{Turner_and_Stone_2001}.

We want to freeze the evolution of any other quantity but the temperature (and, therefore, the gas internal energy).
To do so, first, we build our test model as a static and uniform fluid that is initially out of radiative equilibrium, and we suppress external gravity.
Second, we assume the radiation energy $E$ to be large enough compared to the gas internal energy, so that, during the energy exchange between matter and radiation throughout
the evolution process, $E$ can be considered as a constant quantity. In this respect, the radiation flux, which is related to the gradient of radiation energy
through the FLD relation (\ref{eq:FLD_Fick_law}), is negligible, and the continuity, matter momentum, and radiation energy equations
(\ref{eq:Mass_c}, \ref{eq:Momentum_c}, \ref{eq:Er_conservation}) become irrelevant. The only remaining relevant equation to be solved is the matter energy equation
(\ref{eq:Plasma_energy_c}), but in the form (\ref{eq:system_code_b}).

Even though we use our non-LTE discretization scheme with the radiation emission (or radiative losses) quantity $L$ in the equations (\S~\ref{subsec:substep_2}),
we set $L$ to be equal to the LTE emission
($L = k_P \, \rho\, c\, a_{R} \, T^{4}$). This way, Eq.~(\ref{eq:system_code_b}) can be recast as the following ordinary differential equation:
\begin{subequations} 
  \begin{empheq}[left={\empheqlbrace}]{alignat=4}
      &\frac{d\Em}{dt} \, = \, A \, \left[1 - \left( \frac{\Em}{\Em_{\rm f}} \right)^{4} \right] \label{eq:gas_energy_Turner_test} \\
      &\text{with the two following constants:} \nonumber \\
      &A \, = \,k_P \, \rho\, c \, E                         \label{eq:A_cst}   \\
      &\Em_{\rm f} \, = \, \frac{\rho \, \kB}{(\gamma -1) \, \mu \, \mH \, a_{R}^{1/4}} \; E^{1/4}  \label{eq:E_final} 
  \end{empheq}
\end{subequations}
$\Em_{\rm f}$ is the gas internal energy in the final radiative equilibrium state.

\subsubsection{Setup}
For this test, we set the density $\rho = 10^{-7}~$g cm$^{-3}$, the radiation energy density $E = 10^{10}~$erg cm$^{-3}$,
the Planck\footnote{The choice of the Rosseland mean opacity is irrelevant since it is associated to the radiation flux that is negligible in our test;
however, for computational convenience, we also adopted $\rho \, k_R = 4\times10^{-8}~\rm{cm}^{-1}$.}
 mean opacity $\rho \, k_P = 4\times10^{-8}~\rm{cm}^{-1}$,
the mean molecular weight $\mu=0.6$, and the ratio of specific heats $\gamma = 5/3$.
The domain consists of a cuboid that has a length much larger than the other two dimensions.
In particular, the cuboid has a width and a height of 3 cm, and a length of 100 cm.
The grid consists of $3\times3\times100$ points. All the boundaries are periodic. For this test the hydrodynamic solver is turned off.

The simulation starts at $t = 0~$s with an initial time step of $\delta t = 10^{-20}$~s. After each step,
the time step increases by $0.1\%$. The test is performed using 3 different initial gas internal energy densities
$\Em_0 = (6.4\times10^3;\,6.4\times10^7;\,6.4\times10^8)~$erg cm$^{-3}$.

\subsubsection{Results}
\begin{figure}[ht]
    \centering
    \includegraphics[scale =0.5]{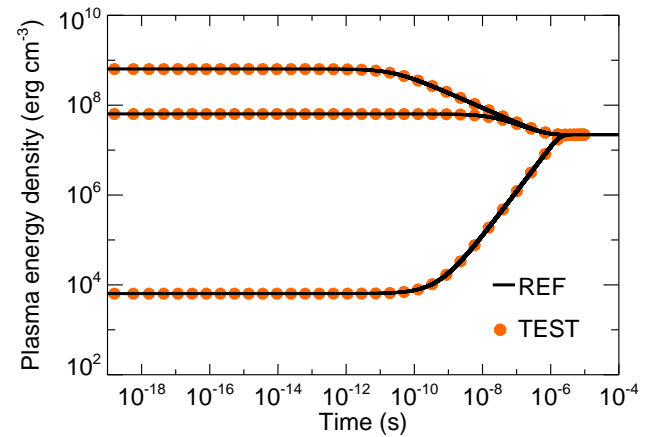}
    \caption{Coupling test with three different initial gas internal energy densities.
             The black line represents the reference solution, the orange dots are the results obtained with the radiation module.}
    \label{fig:coupling_rm}
\end{figure}
Fig. \ref{fig:coupling_rm} shows the comparison between the solution found with our model (red dots) and the semi-analytical reference one
(black line) obtained by solving (\ref{eq:gas_energy_Turner_test}) with a fourth order Runge-Kutta scheme (alternatively,
a full analytical solution is  provided by \citealt{Swesty_and_Myra_2009}),
for the three different initial energy radiation densities $\Em_0$.
The agreement between the two solutions is excellent for all the three cases. 

\subsection{Radiative shocks} \label{subsec:radiative_shocks}
We test our implementation of the full set of non-LTE radiation hydrodynamics equations in PLUTO (\ref{eq:Mass_c}-\ref{eq:gas_energy_temperature_c})
(without external gravity), by assessing the ability of the code to reproduce the structure of a radiative shock. 
We simulate a simple shock case in a quasi-one-dimensional domain; this test follows from \cite{Ensman_1994}.
It is possible to compare some characteristic quantities derived from the test, with analytical estimates \citep{Mihalas_and_Mihalas_book_1984}.
We compare our results with the original version of the code. 

\subsubsection{Setup}
For this test, we simulate an initially uniform fluid with density $\rho = 7.78 \times 10^{-10}~$g cm$^{-3}$, and temperature $T = 10~$K,
moving with velocity $\varv$ along z-axis. We set the ratio of specific heats $\gamma = 7/5$,
and the mean molecular weight $\mu= 1$, in analogy to \cite{Ensman_1994}.
We impose a constant opacity $k_R\times \rho = k_P \times \rho = 3.1\times 10^{-10}~$cm$^{-1}$,
and the initial radiation energy density is set by the equation $E = a_R T^4$.
As in the previous test case (\S~\ref{subsec:radiation_matter_coupling}), we impose the radiative losses $L = k_P \, \rho\, c\, a_{R} \, T^{4}$ (LTE radiation emission),
while using our non-LTE discretization scheme with $L$ in the equations. 

The computational domain has a width and a height of $3.418\times 10^7~$cm, and a length (z-axis) of $7\times 10^{10}~$cm.
Along with \cite{Kolb_et_al_2013}, we have chosen a grid composed of $4\times4\times2048$ cells, the Minerbo flux-limiter
(Eq.~\ref{eq:lambda_Minerbo_1978} in \S~\ref{subsubsec:approx_4_FLD}), a Lax-Friedrichs scheme, 0.4 as CFL (Courant-Friedrichs-Lewy) value,
and a relative tolerance $\epsilon = 10^{-5}$ for the matrix solver.
The lateral boundaries are periodic.
In the direction of the fluid flow (z-axis), we use a reflective boundary at the bottom of the domain ($z=0$), to generate the shock,
and a zero-gradient condition at the top ($z=z_{\rm max}=7\times 10^{10}~$cm).

Moving from $z_{\rm max}$ to $z=0$, the fluid impacts onto the reflective boundary, which generates a shock that propagates back into the fluid.
The fluid velocity is calculated with respect to the computational domain, which is the physical frame of reference.
The shock can be subcritical (low velocity) or supercritical (large velocity) \citep{Mihalas_and_Mihalas_book_1984}.
We simulate both cases: we set $\varv = 6\times10^5~$cm s$^{-1}$ to obtain a subcritical shock,
and $\varv= 20\times10^5~$cm s$^{-1}$ to obtain a supercritical shock.

\subsubsection{Results}

\begin{figure}[ht]
    \centering
    \includegraphics[scale =0.5]{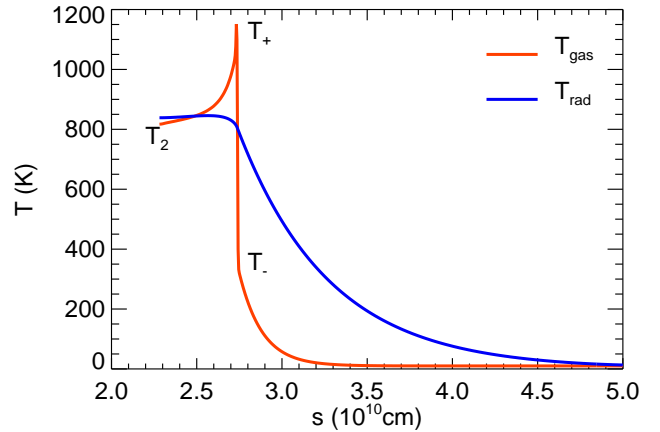}
    \centering
    \includegraphics[scale =0.5]{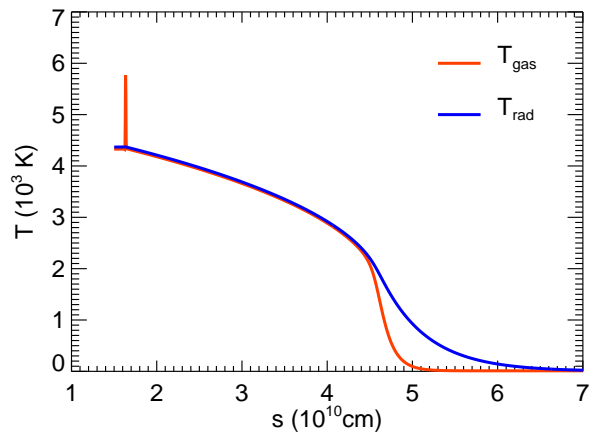}
    \caption{Gas temperature (red) and radiation temperature (blue) versus $s=z-\varv \, t$ for the sub (top panel) and supercritical (bottom panel) shocks. The subcritical shock is shown at time $t=3.8\times 10^4~$s and the supercritical shock at $t=7.5 \times 10^3~$s.}
    \label{fig:radiative_shocks}
\end{figure}

In order to compare our results with those obtained by \citet{Ensman_1994} or \cite{Kolb_et_al_2013}, we introduce the quantity $s=z-\varv \, t$.
Fig.~\ref{fig:radiative_shocks} shows the gas temperature (red) and the radiation temperature (blue), $T_{\rm rad} = \left(E/a_R\right)^{1/4}$, versus $s$
for the subcritical shock (top panel) and the supercritical shock (bottom panel).
In the supercritical shock the temperature after and before the shock are equal, as expected.

In the case of a subcritical shock, some characteristic gas temperatures can be estimated analytically \citep{Mihalas_and_Mihalas_book_1984}:
$T_-$, the temperature immediately ahead of the shock front;
$T_+$, the temperature immediately behind the shock front (Zel'dovich spike: \citealt{Zeldovich_and_Raizer_1967});
$T_2$, the final equilibrium post-shock temperature, reached after radiative cooling.
We have:
\begin{subequations} \label{eq:characteristic_temperatures}
  \begin{empheq}[left={\empheqlbrace}]{alignat=4}
      &T_2 \, & &\approx \, \frac{2\,(\gamma-1\, u^2}{R_G\,(\gamma+1)^2}                                \label{eq:T2_approx} \\
      &T_- \, & &\approx \, \frac{\gamma-1}{\rho \, u \, R_G}\frac{2\,\sigma_{\rm SB}\,T_2^4}{\sqrt{3}} \label{eq:Tm_approx} \\
      &T_+ \, & &\approx \, T_2 \, + \, \frac{3-\gamma}{\gamma+1} \,T_-                                 \label{eq:Tp_approx}
  \end{empheq}
\end{subequations}
where $R_G = k_B/\mu \mH$ is the perfect gas constant, $\sigma_{\rm SB}$ is the Stefan-Boltzmann constant,
and $u$ is the velocity of the shock relative to the upstream fluid.

Table \ref{table:estimate} shows the comparison between our numerical solution, the analytical estimate,
and the solution reported by \cite{Kolb_et_al_2013}. 
Our numerical solution agrees very well with the original version of the code.
The relative deviation with respect to the analytical estimate is no larger than $7\%$.
Moreover, the position and the shape of the shocks are very well reproduced.
We can conclude that our modifications in the radiation module have maintained the accuracy of the code.

\begin{table}
\caption{Comparisons of the characteristic temperatures of a subcritical shock, $T_2$, $T_-$\,, $T_+$\, (Eq.~\ref{eq:characteristic_temperatures}a,\,b,\,c),
obtained from analytical estimates, from our non-LTE code, and from the LTE initial code by \cite{Kolb_et_al_2013}.
The last column shows the relative deviation between our numerical solution and the analytical estimate. \label{table:estimate}}
\centering
\begin{tabular}{l l l l l}
\hline \hline
        & Analytical        & Our numerical & Kolb          & Deviation    \\
        & estimate          &  solution     & et al. (2013) &               \\
\hline
  $T_2$ & $\approx $  865 K &  817 K        &  816.6 K      & $\approx 5\%$ \\
  $T_-$ & $\approx $  315 K &  332 K        &  331.9 K      & $\approx 5\%$ \\
  $T_+$ & $\approx $ 1075 K & 1151 K        & 1147.1 K      & $\approx 7\%$ \\
\hline
\end{tabular}
\end{table}

\section{LTE vs non-LTE radiative shocks} \label{sec:LTE_vs_NLTE_shocks}

The purpose of this section is to show the crucial importance of considering the appropriate regime (LTE vs non-LTE) for given physical conditions,
in order to correctly model the structure and dynamics of a radiating fluid. This is because opacities and radiation emissions
can differ by several orders of magnitudes between both regimes, as exemplified by Fig.~\ref{fig:opacities_NLTE_vs_LTE}.
Such deviations can have a major impact on the momentum and energy exchanges between matter and radiation, and, therefore, on the structure of the flow.

We modify the shock test from \cite{Ensman_1994}, described in \S~\ref{subsec:radiative_shocks},
to have physical conditions quite similar to those in accretion shocks in young stars (\citealt{Sacco_et_al_2008,Colombo_et_al_2019_2}).
As in the two test cases, we use the non-LTE discretization scheme
with the radiation emission (or radiative losses) quantity $L$ in the equations (\S~\ref{subsec:substep_2}).
In one case, referred to as the ``non-LTE case'', we use the non-LTE radiative database (calculated by \citealt{Rodriguez_et_al_2018}: cf. \S~\ref{sec:opacity}),
and let the system evolve following the flow conditions,
for which, at a given time and position, either non-LTE or LTE regime prevails and is self-consistently taken into account by the database.
In the other case, referred to as the ``LTE case'', we use the LTE database (still calculated by \citealt{Rodriguez_et_al_2018}),
and, therefore, force the LTE regime, no matter the physical conditions.

\subsection{Setup} \label{subsec:LTE_vs_NLTE_shocks_setup}
We simulate a uniform fluid with initial density $n = 10^{12}~$cm$^{-3}$, and temperature $T=2\times10^4~$K,
moving with velocity $\varv = 5\times10^7~\rm{cm}^{-1}$ along z-axis. We set $\gamma = 7/5$ and $\mu \approx 1.29$, i.e., assume solar abundances.
Note that here, unlike the preceding test cases, we do not impose a constant value for $k_P$, $k_R$ and $L$, but we use the radiative databases.
The initial radiation energy $E$ is chosen in order to start with a fluid in radiative equilibrium.

The computational domain describes a box of length $10^8~$cm in the non-LTE case, and $10^7~$cm in the LTE case,
and of equal width and height, $3.418\times10^7~$cm in both cases. The grid is composed of 3x3x1024 cells.

We adopt the Minerbo flux-limiter, and boundary conditions identical to those in the shock tests (\S~\ref{subsec:radiative_shocks}).
For the CFL condition, we use 0.01.
For the matrix solver we set a relative tolerance $\epsilon = 10^{-4}$. 

\subsection{Results} \label{subsec:LTE_vs_NLTE_shocks_results}
\begin{figure}
    \centering
    \includegraphics[scale = 0.45]{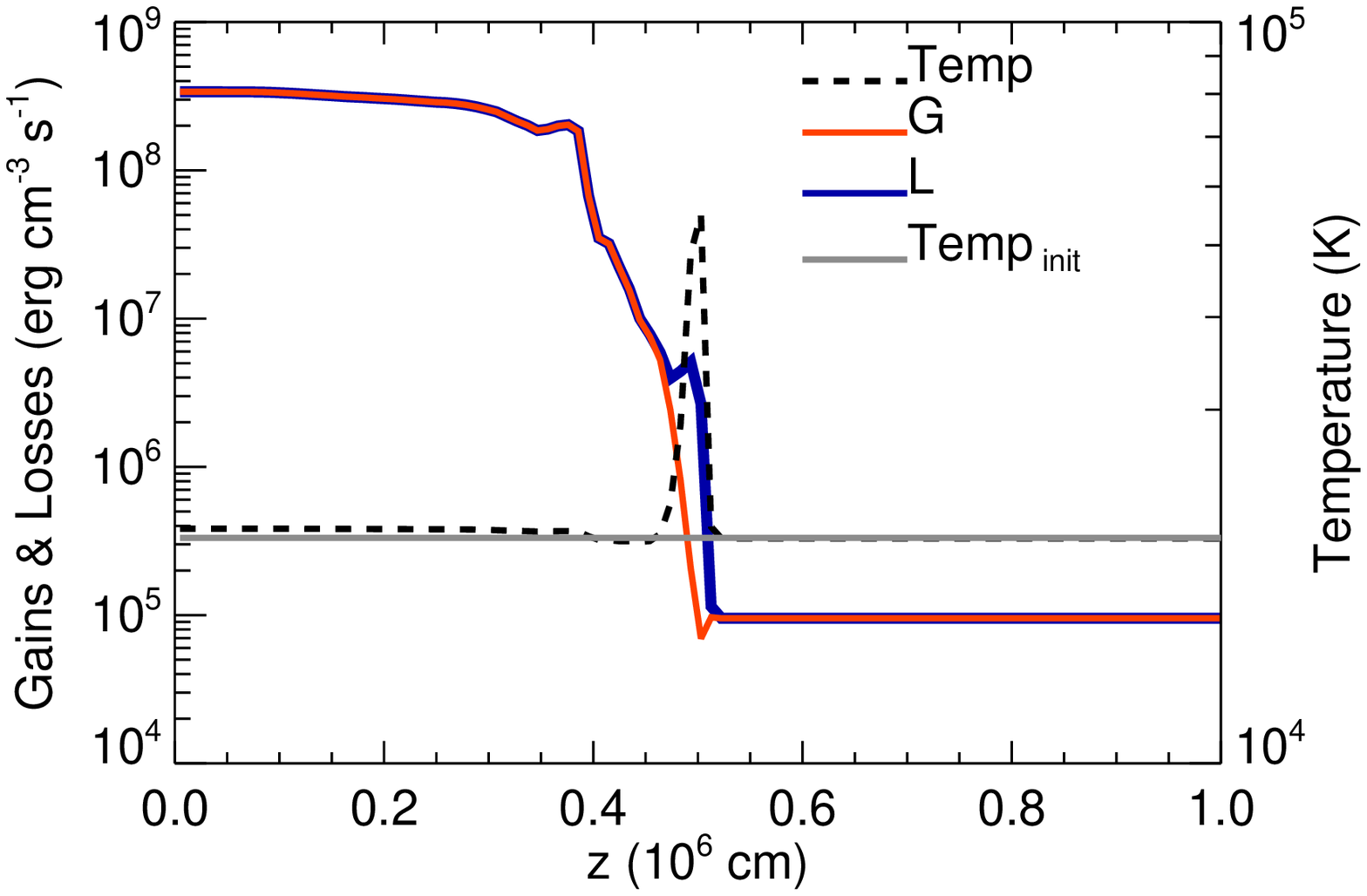}\\
    \includegraphics[scale = 0.45]{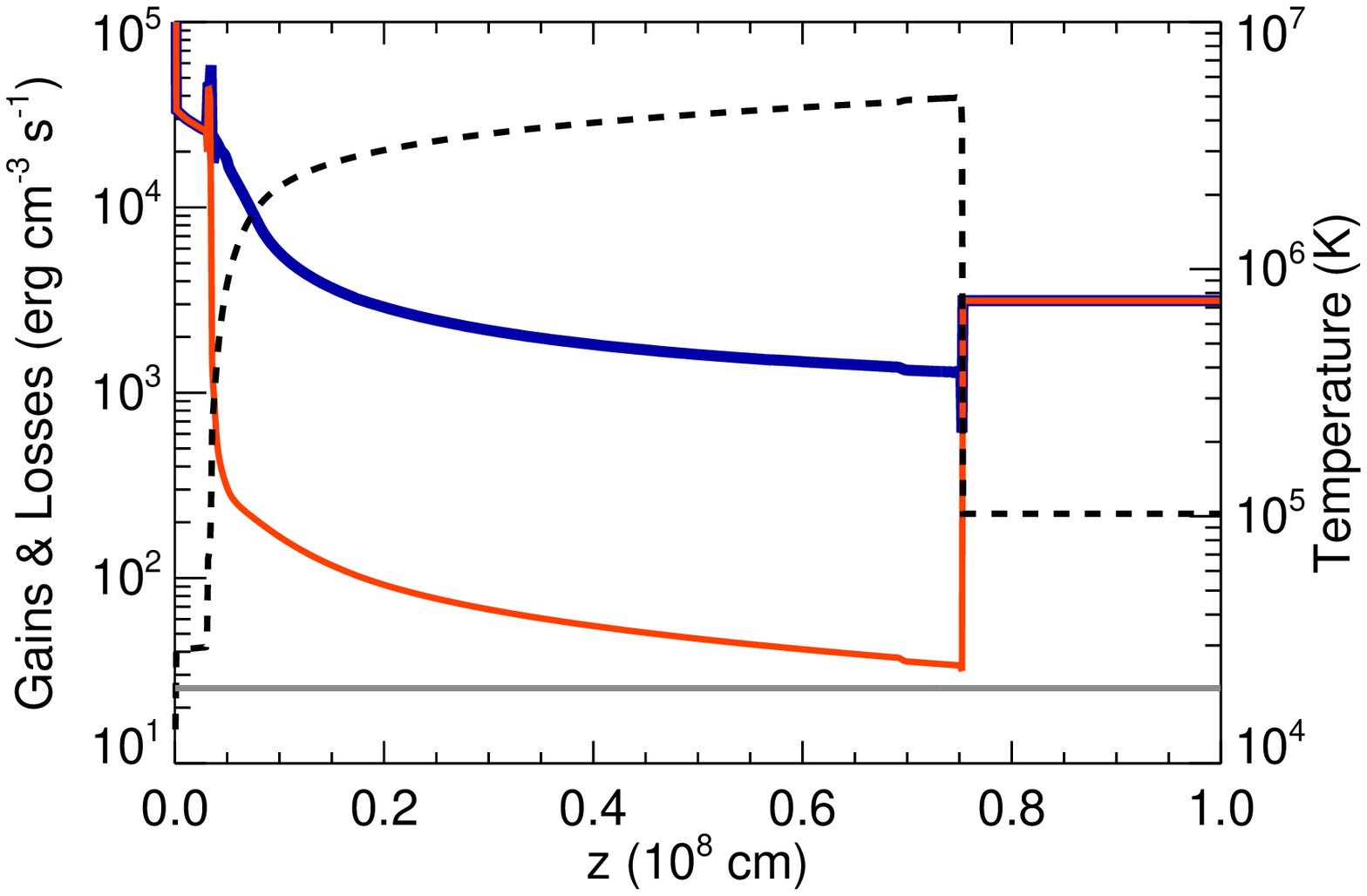}
    \caption{Gas temperature (black dashed line), radiative gains $G$ (red line) and radiative losses $L$ (blue line) versus $z$ for
             the LTE (top panel) and non-LTE (bottom panel) shock cases. Both cases are shown after 6~s of evolution.
             In each plot the grey solid line represents the initial temperature in the domain.}
    \label{fig:LTEvsNLTE}
\end{figure}

The fluid impacts onto the reflective boundary, which generates a shock that propagates back into the fluid.
The shock heats up the material forming a post-shock region.
Fig.~\ref{fig:LTEvsNLTE} shows the profiles of temperature, radiative gains ($G =k_P \, \rho \, c \, E$), which represent the energy gained by the fluid
after absorbing radiation, and radiative losses ($L$) after 6~s of evolution.

The two cases present several differences while we follow their evolutions. This is related to the different regimes taken into account. 
In the LTE case (Fig.\ref{fig:LTEvsNLTE}, top panel), the shock reaches a maximum temperature of $\sim 5\times 10^4~$K (dashed line curve).
At this peak, the radiative losses (blue curve: $L \sim 5 \times 10^6~{\rm erg}\;{\rm cm^{-3}}\;{\rm s^{-1}}$)
are higher than the radiative gains (red curve: $G \sim 5 \times 10^5~{\rm erg}\;{\rm cm^{-3}}\;{\rm s^{-1}}$).
As a consequence, the material rapidly cools down until radiative equilibrium ($G = L$),
and the post-shock region remains relatively cold at $\sim 2\times 10^4~$K, close to the initial temperature.

Even though the radiative losses are extremely high compared to the non-LTE case (see below, the last paragraph of this section), there is no precursor region.
We can invoke two reasons for this:
first, the region that emits is quite small, so the radiation energy emitted per unit time is not enough to heat up the unshocked plasma;
second, according to Fig.\ref{fig:opacities_NLTE_vs_LTE} (mid-right panel),
the Planck opacity, $k_P$, in LTE regime is smaller by several order of magnitudes compared to that in the non-LTE case.
Therefore, matter absorbs far less radiation in LTE than in non-LTE
(a reminder: the gain of radiation energy by matter is $G =k_P \, \rho \, c \, E$).

In the non-LTE regime (Fig.\ref{fig:LTEvsNLTE}, bottom panel), the radiative losses are, in the shocked region,
smaller by around three orders of magnitude compared to those in the LTE regime:
$L_{\,\rm LTE }\sim 5 \times 10^6~{\rm erg}\;{\rm cm^{-3}}\;{\rm s^{-1}}$ (see above),
$L_{\,\rm NLTE }\sim 10^3 - 10^4~{\rm erg}\;{\rm cm^{-3}}\;{\rm s^{-1}}$.
In this case, the shock heats the material up $T \sim 5 \times 10^6~$K,
and generates a hot post-shock region.
Since the radiative losses are smaller than in the LTE case, the shock-heated material needs more time to cool down, and forms a hot slab. 
After $6~$s, the radiative losses trigger the thermal instabilities at the base of the post-shock:
the material rapidly loses thermal energy through radiation emission.
This drop in temperature produces enough radiation energy to heat up the unshocked plasma,
thereby generating a precursor region with a temperature of $T \sim 10^5~$K (dashed line curve on the right in Fig.\ref{fig:LTEvsNLTE}).

\section{Conclusions} \label{sec:conclusion}

Including the effects of radiation in HD and MHD models is a mandatory task to fully describe many astrophysical systems.
Several codes fulfill this request, but none of them consider the more general non-LTE regime.

Here, we have presented our extended version of the LTE radiation module developed by \cite{Kolb_et_al_2013} and implemented in the PLUTO code.
The upgraded module is now able to handle non-LTE regimes (including, self-consistently, the particular case of LTE regime,
depending on the physical plasma conditions).
We use an operator-split method. The system is solved in two substeps, an explicit one for the hyperbolic subsystem, and an implicit one for the subsystem
that involves radiation diffusion and radiation source terms.
It is this second subsystem that we have upgraded so that it can now handle non-LTE conditions.

Starting from the general frequency-integrated comoving-frame radiation hydrodynamics equations, we have reviewed all the assumptions and approximations that
have led to the equations that are actually coded in PLUTO. In particular, we use the flux-limited diffusion approximation.
Also, our implementation is valid for plasma in conditions ranging from free streaming regime to static diffusion regime.
It cannot describe the dynamic diffusion regime: to do so, we have to include the two advection terms in the radiation energy equation.
Moreover, the multigroup implementation in non-LTE can be a further improvement of our module.
These possible developments will be the subject of future works.

The module needs, as input data versus density and temperature, the following radiative quantities:
the radiation emission, the Planck mean opacity, and the Rosseland mean opacity.
Our module currently uses non-LTE databases generated with a CRSS model for a plasma with solar-like abundances.
But the user may provide any other set of databases that would be more appropriate to the problem to be investigated.
In our CRSS approach, the absorption processes, such as photoabsorption and photoionization, are ignored in the determination
of the atomic level populations. Considering such processes is a much more difficult problem that leads to stiff rate equations and stiff coupling with
the thermodynamic state. 

We have tested the new implementation of the non-LTE equations, but in LTE conditions, and have compared our results with semi-analytical solutions,
and with the results given by the previous version of the code.
These tests have established the validity of our implementation.

We also have demonstrated, through the comparison of the structure of a radiative shock, in LTE and in non-LTE regimes, the importance of considering the appropriate regime in order to correctly describe the dynamics of a radiating fluid.

We have already successfully applied this new upgraded version of PLUTO to demonstrate the existence of a radiative precursor in the accreting stream
onto the surface of a classical T Tauri Star \citep{Colombo_et_al_2019_2}.

The radiation module has been implemented in version 4.0 of PLUTO. The code is available, under request, on the observatory of Palermo website\footnote{\url{http://cerere.astropa.unipa.it/progetti_ricerca/HPC/resources.htm}}, to the scientific community.

\begin{acknowledgements}
The authors are grateful to Dr. Mario Flock for useful discussions
on the implementation of radiation equations in the MHD PLUTO code.
PLUTO is developed at the Turin Astronomical Observatory in
collaboration with the Department of Physics of Turin University.
We acknowledge the "Accordo Quadro INAF-CINECA (2017)” the CINECA
Award HP10B1GLGV and the HPC facility (SCAN) of the INAF – Osservatorio Astronomico di Palermo, for the availability of high performance computing resources and support.  This work was supported by the Programme National de Physique Stellaire (PNPS) of CNRS/INSU co-funded by CEA and CNES. This work has been done within the LABEX Plas@par project, and received financial state aid managed by the Agence Nationale de la Recherche (ANR), as part of the programme "Investissements d'avenir" under the reference ANR-11-IDEX-0004-02.
\end{acknowledgements}

\bibliographystyle{aa} 
\bibliography{biblio}

\end{document}